 \documentclass[aps, onecolumn, showpacs, amsmath]{revtex4-2}
\usepackage{dcolumn}
\usepackage{bm}
\usepackage{graphicx}
\usepackage{color}
\usepackage{hyperref}
\usepackage{amsthm}
\usepackage{amssymb}

\begin{document}
\def\be{\begin{equation}}
\def\ee{\end{equation}}

\def\bc{\begin{center}}
\def\ec{\end{center}}
\def\bea{\begin{eqnarray}}
\def\eea{\end{eqnarray}}
\newcommand{\avg}[1]{\langle{#1}\rangle}
\newcommand{\Avg}[1]{\left\langle{#1}\right\rangle}

\def\ie{\textit{i.e.}}
\def\etal{\textit{et al.}}
\def\m{\vec{m}}
\def\G{\mathcal{G}}

\newcommand{\davide}[1]{{\bf\color{blue}#1}}
\newcommand{\gin}[1]{{\bf\color{green}#1}}

\title{Extremal statistics for a one-dimensional Brownian motion with a reflective boundary}
\author{Feng Huang$^1$}
\author{Hanshuang Chen$^2$}\email{chenhshf@ahu.edu.cn}
\affiliation{$^1$School of Mathematics and Physics \& Key Laboratory of Architectural Acoustic Environment of Anhui Higher Education Institutes, Anhui Jianzhu University, Hefei 230601, China \\ $^2$School of Physics and Optoelectronic Engineering, Anhui University, Hefei 230601, China}
\begin{abstract}
In this work, we investigate the extreme value statistics of a one-dimensional Brownian motion (with the diffusion constant $D$) during a time interval $\left[0, t \right]$ in the presence of a reflective boundary at the origin, when starting from a positive position $x_0$. We first obtain the distribution $P(M|x_0,t)$ of the maximum displacement $M$ and its expectation $\langle M \rangle$. In the short-time limit, i.e., $t \ll t_d$ where $t_d=x_0^2/D$ is the diffusion time from the starting position $x_0$ to the reflective boundary at the origin, the particle behaves like a free Brownian motion without any boundaries. In the long-time limit, $t \gg t_d$, $\langle M \rangle$ grows with $t$ as $\langle M \rangle \sim \sqrt{t}$, which is similar to the free Brownian motion, but the prefactor is $\pi/2$ times of the free Brownian motion, embodying the effect of the reflective boundary. By solving the propagator and using a path decomposition technique, we then obtain the joint distribution $P(M,t_m|x_0,t)$ of $M$ and the time $t_m$ at which this maximum is achieved, from which the marginal distribution $P(t_m|x_0,t)$ of $t_m$ is also obtained. For $t \ll t_d$, $P(t_m|x_0,t)$ looks like a U-shaped attributed to the arcsine law of free Brownian motion. For $t$ equal to or larger than order of magnitude of $t_m$, $P(t_m|x_0,t)$ deviates from the U-shaped distribution and becomes asymmetric with respect to $t/2$. Moreover, we compute
the expectation $\langle t_m \rangle$ of $t_m$, and find that $\langle t_m \rangle/t$ is an increasing function of $t$. In two limiting cases, $\langle t_m \rangle/t \to 1/2$ for $t \ll t_d$ and $\langle t_m \rangle/t  \to (1+2G)/4 \approx 0.708$ for $t \gg t_d$, where $G\approx0.916$ is the Catalan's constant. Finally, we analytically compute the statistics of the last time $t_\ell$ the particle crosses the starting position $x_0$ and the occupation time $t_o$ spent above $x_0$. We find that $\langle t_\ell \rangle/t \to 1/2$ in the short-time and long-time limits, and reaches its maximum at an intermediate value of $t$. The fraction of the occupation time $\langle t_o \rangle/t$ is a monotonic function of $t$, and tends towards 1 in the long-time limit.  All the theoretical results are validated by numerical simulations.  
\end{abstract}
\maketitle
\section{Introduction}\label{sec1}
Despite infrequent occurrences, extreme events are of vital importance as they may bring devastating consequences. Examples include natural calamities, such as earthquake, tsunamis and floods, economic collapses, and outbreak of pandemic  \cite{fisher1928limiting,gumbel1958statistics,leadbetter2012extremes,bouchaud1997universality,davison2015statistics,albeverio2006extreme}. 
Extreme-value statistics (EVS) has been a branch of statistics which deals with the extreme deviations of a random process from its mean behavior. For independent and identically distributed random variables, it is known that the extreme-value distribution falls into three famous universality classes,
namely, Gumbel, Fr\'echet, and Weibull depending on the
tails of the distribution of random variables \cite{gnedenko1943distribution}. The study of EVS has also become extremely important in the field of disordered
systems \cite{fyodorov2008freezing,fyodorov2010multifractality}, fluctuating interfaces \cite{raychaudhuri2001maximal,majumdar2004exact}, interacting
spin systems \cite{bar2016exact}, stochastic transport models \cite{majumdar2010real,guillet2020extreme}, random matrices \cite{dean2006large,majumdar2009large,majumdar2014top}, epidemic outbreak \cite{dumonteil2013spatial}, and computer search algorithms \cite{krapivsky2000traveling,majumdar2002extreme,majumdar2003traveling}.

In recent years, there is an increasing interest in studying the EVS for strongly correlated stochastic processes \cite{majumdar2010universal,schehr2014exact,lacroix2020universal,PhysRevLett.111.240601,PhysRevLett.117.080601,PhysRevLett.129.094101,PhysRevLett.130.207101}. We refer the reader to \cite{fortin2015applications,majumdar2020extreme} for two recent reviews on the subject. One of the central goals on this subject is to compute
the statistics of extremes, i.e., the maximum $M$ of a given
trajectory $x(t)$ during an observation time window $\left[0, t \right] $, and the time $t_m$ at which the maximum $M$ is reached. A paradigmatic example is a one-dimensional Brownian motion for a fixed duration $t$ starting from the origin, where the joint distribution of $M$ and $t_m$ is given by \cite{schehr2010extreme}
\begin{eqnarray}\label{eq0.1}
P_0(M, t_m|t)=\frac{M}{{2\pi D\sqrt {t_m^3\left( {t - {t_m}} \right)} }}{e^{ - {M^2}/4Dt_m}}
\end{eqnarray}
with the diffusion constant $D$. Marginalization of Eq.(\ref{eq0.1}) over $t_m$, one gets the one-sided Gaussian distribution of $M$,  
\begin{eqnarray}\label{eq0.2}
P_0(M|t)= \frac{\Theta(M)}{\sqrt{\pi D t}} e^{-M^2/4Dt}, 
\end{eqnarray}
where $\Theta(z)$ is the Heaviside step function such that $\Theta(z)=1$ if $z>0$ and $\Theta(z)=0$ otherwise. In particular, from Eq.(\ref{eq0.2}) one can easily obtain that the expected maximum $M(t)$ of a free Brownian motion grows like $\langle M(t) \rangle =2\sqrt{D t/\pi}$. On the other hand, one can obtain the marginal distribution of $t_m$ by integrating $P_0(M,t_m|t)$ over $M$,   
\begin{eqnarray}\label{eq0.3}
P_0(t_m|t)= \frac{1}{\pi \sqrt{t_m(t-t_m)}}, \quad 0\leq t_m \leq t,
\end{eqnarray}
which is often referred to as the ``arcsine law" due to P. L\'evy \cite{Levy1940ArcsineLaw,feller1971introduction,majumdar2007brownian}. The name stems from the fact that the cumulative distribution of $t_m$ reads $F(z) =  \int_0^z {P_0(t_m |t)dt_m}  = (2/\pi)\arcsin \sqrt {z/t}$. A counterintuitive aspect of the U-shaped distribution of Eq.(\ref{eq0.3}) is that its average value $\langle t_m \rangle=t/2$ corresponds to the minimum of the
distribution, i.e., the less probable outcome, whereas values
close to the extrema $t_m=0$ and $t_m=t$ are much more likely. In fact, the other two times, i.e., the last time $t_\ell$ the process crosses the origin and the occupation time $t_o$ spent on the positive (or the negative) semi axis, satisfy the same distribution as Eq.(\ref{eq0.3}) as well. The three arcsine laws for Brownian motion play a central role in EVS.

Extreme observables were also studied for variants of Brownian motion.   For a class of constrained Brownian motions including Brownian excursions, Brownian meanders, and reflected Brownian bridge, the joint distribution of the maximum $M$ and the extreme time $t_m$, and their marginal distributions have been analytically obtained \cite{majumdar2008time}. For the random acceleration process, which is one of the simplest non-Markov stochastic processes, the distribution of $t_m$ \cite{majumdar2010time} and the first two moments of the occupation time $t_o$ \cite{boutcheng2016occupation} were studied. For the run-and-tumble motion, a representative model of active particles like \textit{E. coli}, the distributions of three times, i.e., $t_m$, $t_\ell$ and $t_o$ were derived in one dimension \cite{SinghArcsinelaws_RTP}, and the results were used to compute the statistical properties of the convex hull of a planar run-and-tumble motion \cite{hartmann2020convex,singh2022mean}. In Refs.\cite{sadhu2018generalized,sadhu2021functionals}, the three arcsine laws were generalized to fractional Brownian motion, which is a non-Markovian Gaussian process indexed by the Hurst exponent $H \in \left(0, 1\right)$, generalizing standard Brownian motion ($H=\frac{1}{2}$) to account for anomalous diffusion. Using a perturbative expansion in $\epsilon=H-\frac{1}{2}$, the distributions of three times were analytically obtained up to second-order. The distributions of $t_m$ and $t_o$ are symmetric with respect to half of the duration time $t/2$. In the leading term, the two distributions are the same, but become distinguishable in the sub-leading term. The distribution of $t_\ell$ is markedly differently from the other two distributions; especially, it is asymmetric with respect to $t/2$. Recently, Brownian motion and fractional Brownian motion subject to stochastic resetting have received increasing attention due to the emergence of a nonequilibrium steady state and its advantage in accelerating a random search \cite{evans2011diffusion,evans2020stochastic,PhysRevE.104.024105,PhysRevE.105.L012106}. For a one-dimensional resetting Brownian motion with a fixed duration, the distributions of  $M$ and $t_m$ have been derived \cite{PhysRevE.103.052119,PhysRevE.103.022135}, from which statistical properties of the convex hull of a planar resetting Brownian motion were obtained \cite{PhysRevE.103.022135}. The distribution of $t_o$ for a one-dimensional resetting Brownian motion is considered, and particular attention was paid to the large deviation property of the distribution \cite{den2019properties}. Very recently, extreme statistics and spacing distribution for the positions of a one-dimensional $N$ Brownian particles under the simultaneous resetting was investigated \cite{PhysRevLett.130.207101}.  EVS has also been studied for general stochastic processes \cite{lamperti1958occupation,kasahara1977limit,dhar1999residence,majumdar2002exact,schehr2010extreme,PhysRevLett.107.170601,PhysRevE.83.061146,PhysRevE.105.024113,PhysRevE.102.032103}. Extension to study the distribution of the time difference
between the minimum and the maximum for stochastic processes has also been made in \cite{mori2019time,mori2020distribution} and to study the EVS before a first passage time through a specified threshold \cite{randon2007distribution,krapivsky2010maximum,singh2022extremal,PhysRevX.7.011019,PhysRevE.108.044115}. The distribution of a related observable, i.e., the number of distinct sites visited by a one-dimensional random walker before hitting a target \cite{PhysRevE.103.032107}, and the joint distribution of the first-passage time to a target and the number of distinct sites visited \cite{PhysRevE.105.034116} have been obtained analytically for several representative Markovian processes, such as simple symmetric random walks, biased random walks, persistent random walks, and resetting random walks. Quite remarkably, the statistics of $t_m$ has found applications in convex hull problems \cite{hartmann2020convex,singh2022mean,randon2009convex,PhysRevE.103.022135} and also in detecting whether a stationary process is equilibrium or not \cite{mori2021distribution,mori2022time}.

So far, most of previous studies were paid their attentions to the EVS of the stochastic process without confinement. However, in many practical situations, the stochastic process takes place in a confined geometry. For example, the range of animal foraging can get constrained by natural or human-built obstacles, such as a river, mountains, urban areas, roads, and so on. A planar Brownian motion in the presence of an infinite reflective wall was studied \cite{chupeau2015convex,chupeau2015mean}. It was shown that the presence of the wall breaks the isotropy of the process and induces a non-trivial effect on the convex hull of the Brownian motion. Very recently, a Brownian particle in a $d$-dimensional ball with radius $R$ with reflecting boundaries was considered, and the statistics of the maximum displacement $M_x(t)$ of along the $x$-direction at time $t$ was studied \cite{de2022statistics}. It was shown the distribution of the fluctuation $(M_x(t)-R)/R$ in the long-time limit exhibits a rich variety of behaviors depending on the dimension $d$. In a very recent work \cite{kay2023extreme}, the authors studied that EVS of a one-dimensional Brownian motion in the presence of a permeable barrier when starting a position infinitely approaching the barrier. They derived the joint distribution of the maximum displacement $M(t)$ and the time $t_m$ at the maximum is reached, from which the marginal distributions of $M(t)$ and $t_m$ were also obtained. It was also reported that the confinement produces a profound impact on the first-passage properties of the fractional Brownian motion \cite{PhysRevE.99.032106} and the ergodicity of heterogeneous diffusion \cite{cherstvy2014ageing}.

In this work, we aim to study the EVS of the Brownian motion in one dimension subject to a reflective boundary at the origin when starting from a positive position $x_0$, such that the Brownian motion is confined in the positive half-space. From an intuitive perspective of view, when the total time duration $t$ is much less than the diffusion time $t_d=x^2_0/D$ from the starting position to the reflective boundary, the reflective boundary has not much impact on the Brownian motion and thus the EVS is similar to that of a free Brownian motion. We should emphasize that $t_d$ is not confused with the mean first-passage time from $x_0$ to the origin. It is well-known that the first-passage time $t_f$ follows $P(t_f) =\frac{1}{\sqrt{4 \pi D t_f^3}} \exp\left( {-{x_0^2}/{4Dt_f}} \right) $. The heavy tail of the distribution leads to the divergence of mean first-passage time. Instead, $t_d$ can be understood as the typical value of the first-passage time since one can easily check that $P(t_f)$ attains its unique maximum at $t_f=t_d$.  A special focus should be paid to the case of $t$ equal to or larger than the order of the magnitude of $t_d$ where the reflective boundary comes into effect. We first derive the survival probability of the Brownian particle in the presence of a reflective boundary at the origin and an absorbing boundary at $x=M$, thus leading to the distribution $P(M|x_0,t)$ of the maximum displacement $M$ within a duration $t$. We also obtain the expectation $\langle M \rangle$ of $M$ as a function of $x_0$ and $t$, from which we find that in the long-time limit $\langle M \rangle \sim \sqrt{\pi D t}$, up to a correction proportional to $x_0^2/\sqrt{t}$. This is qualitatively the same as the case of the free Brownian motion ($\langle M \rangle \sim 2 \sqrt{ D t/\pi}$ for the latter), but with a different prefactor. Moreover, we use a path decomposition technique to obtain the joint distribution $P(M,t_m|x_0,t)$ of the maximum displacement $M$ and the time $t_m$ at which $M$ is reached. Marginalization of $P(M,t_m|x_0,t)$ over $M$ to obtain the distribution of $P(t_m|x_0,t)$. For $t>t_d$, the distribution of $t_m$ deviates from the so-called arcsine law, and is no longer symmetric with respect to $t/2$. Interestingly, we find that the ratio of the expectation of $t_m$ to $t$,  $\langle t_m \rangle/t$, increases monotonically with $t$, and its value approaches 1/2 in the limit $t \ll t_d$ and tends towards another constant $(1+2G)/4 \approx 0.708$ in the opposite limit $t \gg t_d$, where $G$ is the Catalan's constant. At last, we show how the statistics of the other two times, i.e., the last time $t_\ell$ the process passes through the initial position $x_0$ and the occupation time $t_o$ spent above $x_0$, is modified in the presence of the reflective boundary. The expectations of $t_\ell$ and $t_o$ are analytically obtained.  In the short-time and the long-time limits, $\langle t_\ell \rangle/t$ tends towards $1/2$.  Interestingly, $\langle t_\ell \rangle/t$ shows a nonmonotonic dependence on $t$, and a maximum $\langle t_\ell \rangle/t$ occurs at a moderate value of $t$. The fraction of the occupation time $\langle t_o \rangle/t$ varies monotonically from $1/2$ and approaches to $1$ in the long-time limit.

The rest of the paper is organized as follows. In Sec.\ref{sec2} we define the Brownian motion in one dimension subject to a reflective boundary and propose the extreme-value questions we want to study. In Sec.\ref{sec3} and Sec.\ref{sec4}, we derive, both in the time domain and in the Laplace domain, the propagator and survival probability of the Brownian motion in an interval with a reflective end and an absorbing end. In Sec.\ref{sec5}, we present the marginal distribution of the maximum displacement $M$ and its expected value. Using a path decomposition technique, in Sec.\ref{sec6} we obtain the joint distribution of $M$ and the time $t_m$ at which the maximum $M$ is reached. In Sec.\ref{sec7}, we provide the marginal of the extreme time $t_m$ and make asymptotic analysis for the expectation of $t_m$ in the short-time and the long-time limits. In Sec.\ref{sec8}, the statistics of the last time visited the starting position and the occupation time spent above the starting position are analytically given. Finally, the main conclusions and perspective are addressed in Sec.\ref{sec9}.

\begin{figure}
	\centerline{\includegraphics*[width=0.6\columnwidth]{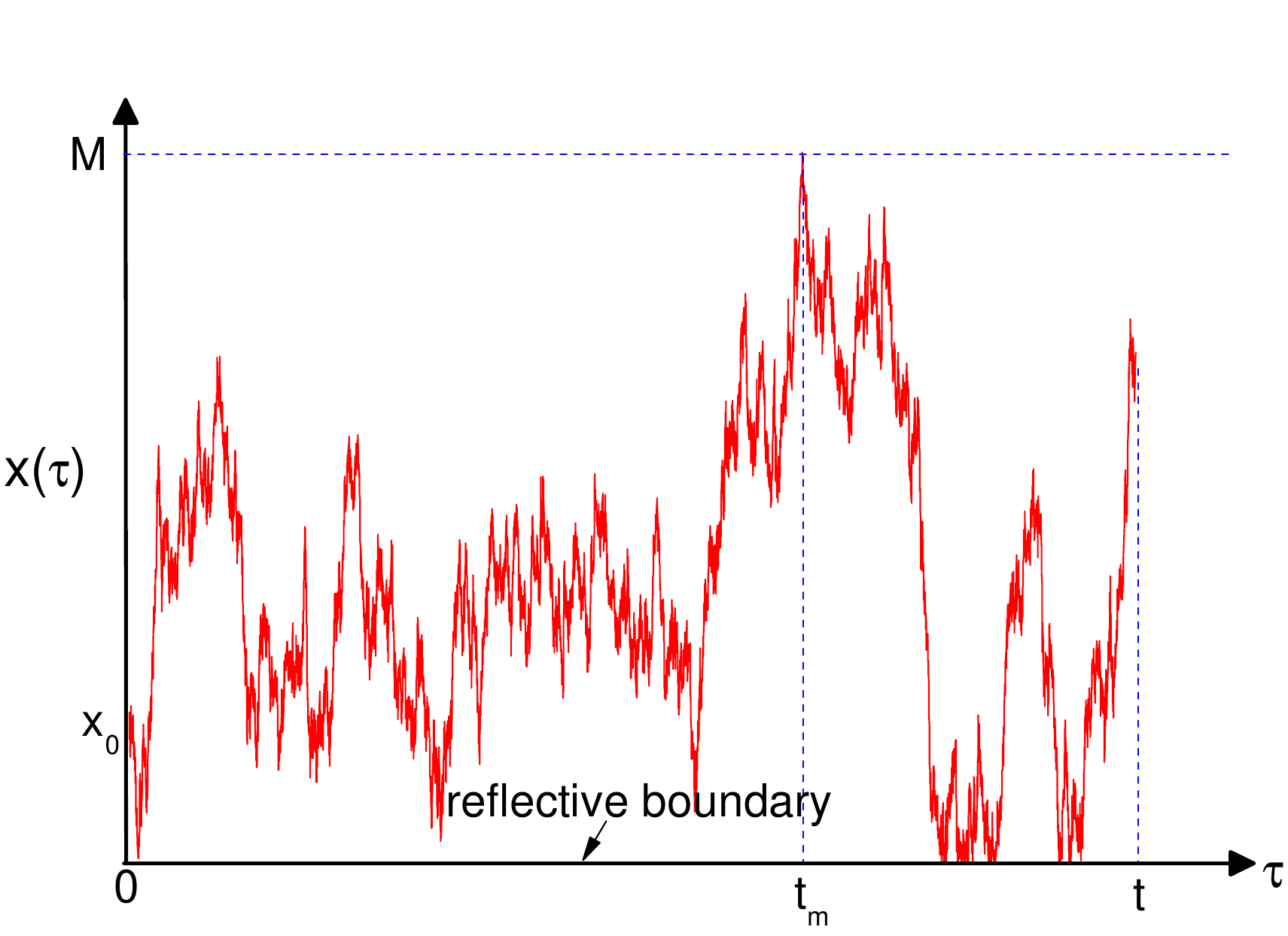}}
	\caption{A realization of a one-dimensional Brownian motion in the presence of a reflective wall at the origin starting from $x_0$. The displacement $x(\tau)$ of the particle reaches its maximum $M$ at the time $t_m$ during a time interval $\left[0, t \right] $. \label{fig1}}
\end{figure}

\section{Model}\label{sec2}
Let us consider a one-dimensional Brownian motion starting at $x_0$($>0$) with a reflective boundary at the origin, 
\begin{eqnarray}\label{eq1.0}
\dot{x}=\sqrt{2D} \xi(t)
\end{eqnarray}
where $D$ is the diffusion constant, and ${\xi }(t)$ is a Gaussian white noise with zero mean $\langle {\xi (t)} \rangle  = 0$ and delta correlator $\langle {\xi (t)}{\xi (t')} \rangle  = \delta ( {t - t'} )$. During a time interval $\left[0, t \right] $, the displacement $x(t)$ of the Brownian reaches its maximum $M$ at time $t_m$, see Fig.\ref{fig1} for an illustration. We are interested in the joint distribution $P(M, t_m|x_0,t)$ of $M$ and $t_m$, and their marginal distributions of $M$ and $t_m$, $P(M|x_0,t)$ and $P(t_m|x_0,t)$.

\section{Propagator with a reflective wall at $x=0$ and an absorbing wall at $x=M$}\label{sec3}
We begin with the computation of the probability density function $G(x,t|x_0)$ of position of the Brownian particle or the propagator in the presence of a reflective wall at $x=0$, when starting from an initial position $x_0 \in \left[0, M\right) $. On the other hand, the maximum displacement $M$ of the particle within a time duration $\left[0, t \right] $ yields an absorbing boundary at $x=M$.

\subsection{Propagator in the time domain}
The Fokker-Planck equation for governing the time evolution of  $G(x,t|x_0)$ reads
\begin{eqnarray}\label{eq2.0}
\frac{{\partial G\left( {x,t|{x_0}} \right)}}{{\partial t}} = D\frac{{{\partial ^2}G\left( {x,t|{x_0}} \right)}}{{\partial {x^2}}}
\end{eqnarray}
Perfect reflection is imposed by requiring that the probability current vanishes at the origin, such that 
\begin{eqnarray}\label{eq2.0.1}
\frac{\partial G(x,t|x_0)}{\partial x}|_{x=0}=0.
\end{eqnarray}	
An absorbing boundary at $x=M$ requires
\begin{eqnarray}\label{eq2.0.2}
G(M,t|x_0)=0.
\end{eqnarray}	
Using separation of variables, Eq.(\ref{eq2.0}) can be solved combined with the boundary conditions Eq.(\ref{eq2.0.1}), Eq.(\ref{eq2.0.2}), and initial condition $G(x,0|x_0)=\delta(x-x_0)$, which gives the series representation of the propagator, 
\begin{eqnarray}\label{eq4.1}
G\left( {x,t|{x_0}} \right) = \frac{2}{M}\sum\limits_{n = 0}^\infty  {\cos \left[ {\frac{{\left( {n + 1/2} \right)\pi {x_0}}}{M}} \right]} \cos \left[ {\frac{{\left( {n + 1/2} \right)\pi x}}{M}} \right] e^{{ - \frac{{{{\left( {n + 1/2} \right)}^2}{\pi ^2}Dt}}{{{M^2}}}} }
\end{eqnarray}

\subsection{Propagator in the Laplace domain} \label{sec3b}
In this subsection, we solve the propagator in the Laplace domain. By the Laplace transformations for Eq.(\ref{eq2.0}), Eq.(\ref{eq2.0.1}) and Eq.(\ref{eq2.0.2}), we obtain 
\begin{eqnarray}
s\tilde G( {x,s|{x_0}} ) - \delta ( {x - {x_0}} ) = D\frac{{{d^2}\tilde G( {x,s|{x_0}} )}}{{d{x^2}}}, \label{eq2.2.1} \\
\tilde G ( {M,s|{x_0}} ) = 0, \quad {\tilde G'} ( {0,s|{x_0}} ) = 0. \label{eq2.2.2}
\end{eqnarray}
Eq.(\ref{eq2.2.1}) can be solved separately in the region $0<x<x_0$ and in the region $x_0<x<M$. In each region the solutions are
\begin{eqnarray}\label{eq2.3}
\tilde G ( {x,s|{x_0}} )= \left\{ \begin{array}{llc}
A_1 e^{\alpha x} + B_1 e^{-\alpha x} ,  & 0<x<x_0,  \\
A_2 e^{\alpha x} + B_2 e^{-\alpha x},   & x_0<x<M,  \\ 
\end{array}  \right. 
\end{eqnarray}
where $\alpha=\sqrt{s/D}$.

The boundary conditions in Eq.(\ref{eq2.2.2}) leads to 
\begin{eqnarray}\label{eq2.4}
\left\{ \begin{gathered}
{A_1}\alpha  - {B_1}\alpha  = 0, \hfill \\
{A_2}{e^{\alpha M}} + {B_2}{e^{ - \alpha M}} = 0. \hfill \\ 
\end{gathered}  \right.
\end{eqnarray}
In addition, we require that $\tilde G ( {x,s|{x_0}} )$ is continuous at $x=x_0$, i.e., 
\begin{eqnarray}\label{eq2.5}
{A_1}{e^{\alpha {x_0}}} + {B_1}{e^{ - \alpha {x_0}}} = {A_2}{e^{\alpha {x_0}}} + {B_2}{e^{ - \alpha {x_0}}}.
\end{eqnarray}
However, due to the presence of the delta function in Eq.(\ref{eq2.2.1}) the derivative of $\tilde G ( {x,s|{x_0}} )$ is not continuous at  $x=x_0$. To find the matching condition for the derivative we integrate Eq.(\ref{eq2.2.1}) from $x_0-\epsilon$ to $x_0 + \epsilon$ (with $\epsilon \to 0^{+}$), which yields
\begin{eqnarray}\label{eq2.6}
{\left. {\frac{{d\tilde G\left( {x,s|{x_0}} \right)}}{{dx}}} \right|_{x = x_0^ + }} - {\left. {\frac{{d\tilde G\left( {x,s|{x_0}} \right)}}{{dx}}} \right|_{x = x_0^ - }} =  - \frac{1}{D}.
\end{eqnarray}
According to Eq.(\ref{eq2.3}), the matching condition in Eq.(\ref{eq2.6}) becomes
\begin{eqnarray}\label{eq2.7}
\left( {{A_2}\alpha {e^{\alpha {x_0}}} - {B_2}\alpha {e^{ - \alpha {x_0}}}} \right) - \left( {{A_1}\alpha {e^{\alpha {x_0}}} - {B_1}\alpha {e^{ - \alpha {x_0}}}} \right) =  - 1/D.
\end{eqnarray}
By solving Eq.(\ref{eq2.4}), Eq.(\ref{eq2.5}) and Eq.(\ref{eq2.7}), the coefficients $A_1$, $B_1$, $A_2$, and $B_2$ are given by
\begin{eqnarray}\label{eq2.7.1}
\left\{ \begin{gathered}
{A_1} = {B_1} = \frac{{\sinh \left[ {\alpha \left( {M - {x_0}} \right)} \right]}}{{2\alpha D\cosh \left( {\alpha M} \right)}} \hfill \\
{A_2} = \frac{{\cosh \left( {\alpha {x_0}} \right)\left[ {\tanh \left( {\alpha M} \right) - 1} \right]}}{{2\alpha D}} \hfill \\
{B_2} = \frac{{\cosh \left( {\alpha {x_0}} \right)\left[ {\tanh \left( {\alpha M} \right) + 1} \right]}}{{2\alpha D}} \hfill \\ 
\end{gathered}  \right.
\end{eqnarray}
Substituting Eq.(\ref{eq2.7.1}) into Eq.(\ref{eq2.3}), we obtain the propagator in the Laplace domain, 
\begin{eqnarray}\label{eq2.8}
\tilde G\left( {x,s|{x_0}} \right) = \frac{{\cosh \left( {\alpha {x_ < }} \right)\sinh \left[ {\alpha \left( {M - {x_ > }} \right)} \right]}}{{\alpha D\cosh \left( {\alpha M} \right)}}
\end{eqnarray}
with $x_<=\min(x,x_0)$ and $x_>=\max(x,x_0)$.

\section{Survival probability with a reflective wall at $x=0$ and an absorbing wall at $x=M$}\label{sec4}
\subsection{Survival probability in the time domain}
It is easy to see that the distribution of the maximum $M$ over the duration $t$ in absence of any absorbing barrier is related to the survival probability of the process $x(\tau)$ in presence of an absorbing barrier at $M$, i.e., $Q(x_0,t|M)={\rm{Prob}}(x(\tau) \leq M | 0\leq \tau \leq t )$, where $Q(x_0,t|M)$ is the survival probability that the process has not yet hit the absorbing boundary at $x=M$ up to time $t$ starting from $x_0$, which is given by \cite{redner2001guide}
\begin{eqnarray}\label{eq4.2}
Q\left( {{x_0},t|M} \right) = \int_0^M dx {G\left( {x,t|{x_0}} \right)}  = \frac{2}{\pi }\sum\limits_{n = 0}^\infty  {\frac{{{{\left( { - 1} \right)}^n}}}{{n + 1/2}}\cos \left[ {\frac{{\left( {n + 1/2} \right)\pi {x_0}}}{M}} \right]} e^{ - \frac{{{{\left( {n + 1/2} \right)}^2}{\pi ^2}Dt}}{{{M^2}}}} 
\end{eqnarray}
\subsection{Survival probability in the Laplace domain}
We will see that in order to obtain the expectation of the maximum displacement $M$ it is convenient to solve survival probability in the Laplace domain. It is known that the survival probability satisfies the backward equation \cite{redner2001guide}, 
\begin{eqnarray}\label{eq1.1}
\frac{{\partial {Q}\left( {{x_0},t|M} \right)}}{{\partial t}} = D\frac{{{\partial ^2}{Q}\left( {{x_0},t|M} \right)}}{{\partial x_0^2}} ,
\end{eqnarray}
subject to the boundary conditions $Q(M,t|M)=0$ and $\partial Q(x_0,t|M)/ \partial x_0|_{x_0=0}=0$ and initial condition $Q(x_0,0|M)=1$ for $x_0 \in \left[ 0, M\right) $. Performing the Lapalce transform, $\tilde{Q}(x_0,s|M)=\int_{0}^{\infty} dt e^{-st} Q(x_0,t|M)$, for Eq.(\ref{eq1.1}) to obtain
\begin{eqnarray}\label{eq1.2}
D\frac{{{d^2}{{\tilde Q}}\left( {{x_0},s|M} \right)}}{{dx_0^2}} -s {{\tilde Q}}\left( {{x_0},s|M} \right)  =  - 1.
\end{eqnarray}
Eq.(\ref{eq1.2}) can be solved subject to the boundary conditions, which yields
\begin{eqnarray}\label{eq1.3}
\tilde Q(x_0,s|M)=\frac{1}{s}-\frac{ \cosh \left(\alpha  x_0 \right) }{s \cosh \left( \alpha M \right)}
\end{eqnarray}
with $\alpha=\sqrt{{s}/{D}}$ again.

\section{Marginal distribution $P(M|x_0,t)$}\label{sec5}
Differentiating Eq.(\ref{eq4.2}) with respect to $M$ gives the probability density function of $M$,
\begin{eqnarray}\label{eq4.3}
P\left( {M|{x_0},t} \right) = \frac{2}{{{M^2}}}\sum\limits_{n = 0}^\infty  {{{\left( { - 1} \right)}^n} e^{ - \frac{{{{\left( {n + 1/2} \right)}^2}{\pi ^2}Dt}}{{{M^2}}}} \left\{ {{x_0}\sin \left[ {\frac{{\left( {n + 1/2} \right)\pi {x_0}}}{M}} \right] + \frac{{2\left( {n + 1/2} \right)\pi Dt}}{M}\cos \left[ {\frac{{\left( {n + 1/2} \right)\pi {x_0}}}{M}} \right]} \right\}} .
\end{eqnarray}

\begin{figure}
	\centerline{\includegraphics*[width=0.6\columnwidth]{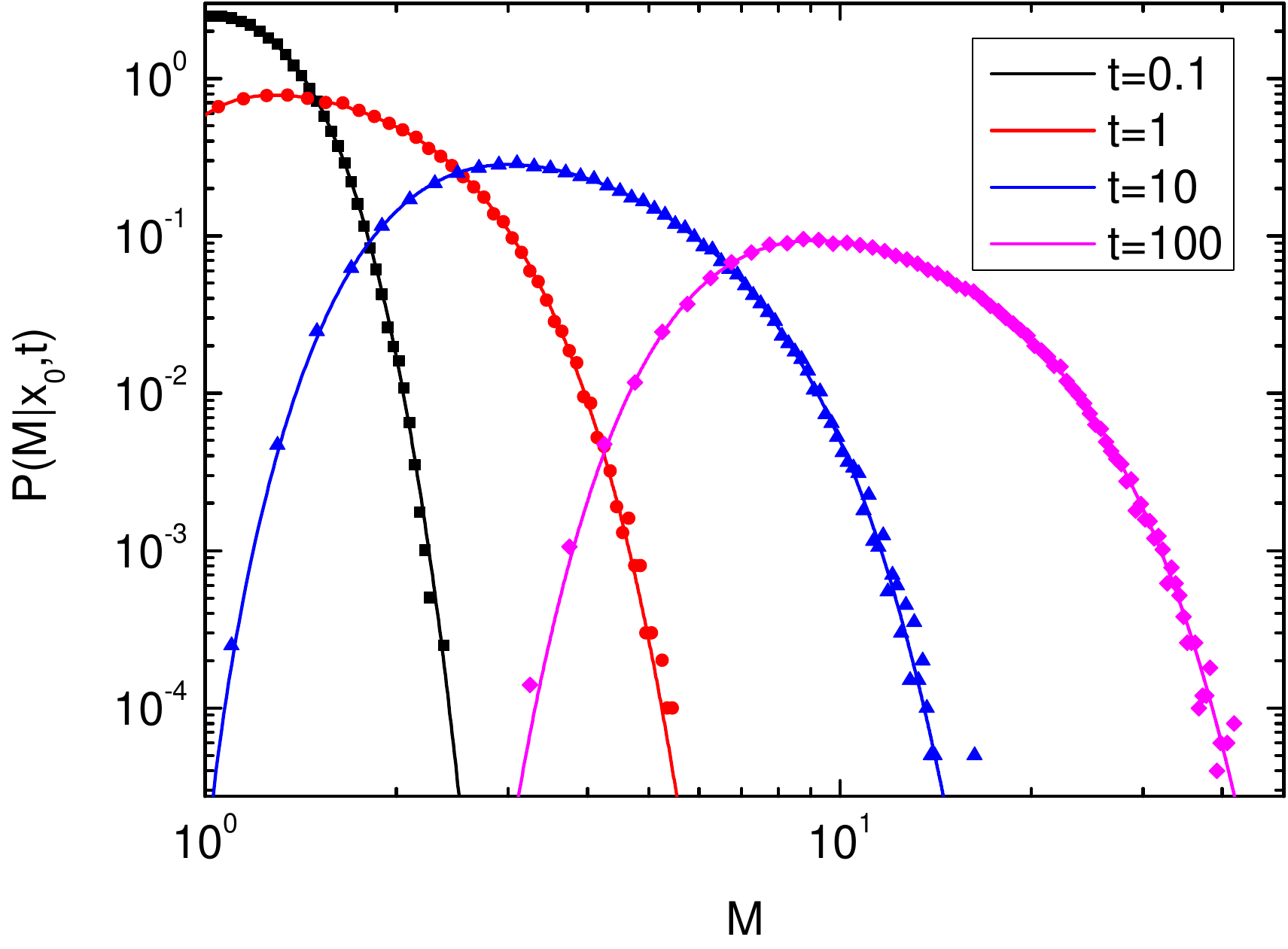}}
	\caption{Marginal distribution $P(M|x_0,t)$ of maximum displacement $M$ for different time duration $t$, where $x_0=1$ and $D=1/2$ are fixed. Line and symbols represent the theory and simulation results, respectively. \label{fig2}}
\end{figure}

In Fig.\ref{fig2}, we plot the marginal distribution $P(M|x_0,t)$ for different $t$, but with fixed $x_0=1$ and $D=1/2$. For comparison, we have performed simulations that supports our theory. In all simulations, we have used the time step $dt=10^{-5}$ and the results are averaged over $10^5$ different realizations. For $t$ much less than the diffusion time $t_d=x_0^2/D$ from the starting position $x_0$ to the reflective wall at the origin, the diffusing particle is almost unaffected by reflective boundary such that the distribution $P(M|x_0,t)$ is similar that of freely diffusive particle shown in Eq.(\ref{eq0.2}), from which we can observe that $P(M|x_0,t)$ decreases monotonically with $M$. As $t$ increases, the reflective boundary comes into effect, and $P(M|x_0,t)$ is no longer a monotonic function of $M$. Interestingly, $P(M|x_0,t)$ can show a unique maximum at an intermediate level of $M$. In the limit $t \gg t_d$, the summation in the right-hand side of Eq.(\ref{eq4.3}) is dominated by only one term for $n=0$. In this case, $P(M|x_0,t)$ decays with $M$ as $M^{-3}$ in the large-$M$ limit.

Taking the derivative of Eq.(\ref{eq1.3}) with respect to $M$, we obtain the marginal distribution $P(M|x_0,t)$ in the Laplace domain,
\begin{eqnarray}\label{eq1.4}
\tilde P(M|x_0,s)=\frac{\partial \tilde Q(x_0,s|M)}{\partial M}  =\frac{\alpha \sinh (\alpha M) \cosh (\alpha x_0)}{s \cosh^2 (\alpha M)}.
\end{eqnarray}
It turns out that inverting Eq.(\ref{eq1.4}) is not a trivial task. However, we can conveniently compute the expectation of the maximum displacement $M$ from $\tilde P(M|x_0,s)$, defined as
\begin{eqnarray}\label{eq1.5}
\langle M(x_0,t) \rangle =\int_{x_0}^{\infty} dM M P(M|x_0,t) .
\end{eqnarray}
Performing the Laplace transform for Eq.(\ref{eq1.5}) and utilizing Eq.(\ref{eq1.4}), we have
\begin{eqnarray}\label{eq1.6}
 \int_{0}^{\infty}   dt e^{-st} \langle M(x_0,t) \rangle = \int_{x_0}^{\infty} dM  M \tilde{P}(M|x_0,s)  = \frac{x_0}{s}+ \frac{2 {\rm {\rm{arctan}}}(e^{-\alpha x_0}) \cosh(\alpha x_0)}{ \alpha s }.
\end{eqnarray}
Prior to the inversion of Eq.(\ref{eq1.6}), we first consider two limiting cases. In the short-time limit ($s \to \infty$) and in the long-time limit ($s \to 0$), Eq.(\ref{eq1.6}) becomes
\begin{eqnarray}\label{eq1.6.1}
\int_{0}^{\infty} dt e^{-st} \langle M(x_0,t) \rangle = \left\{ \begin{array}{cc}
\frac{{{x_0}}}{s} + \frac{{\sqrt D }}{{{s^{3/2}}}},   & s \to \infty,  \\
\frac{{\pi \sqrt D }}{{2{s^{3/2}}}} + \frac{{\pi x_0^2}}{{4\sqrt {Ds} }}, &   s \to 0.  \\ 
\end{array}  \right. 
\end{eqnarray}
Performing the inverse Laplace transform for Eq.(\ref{eq1.6.1}), we obtain
\begin{eqnarray}\label{eq1.6.2}
\langle  M(x_0,t) \rangle= \left\{ \begin{array}{cc}
x_0+\frac{{2\sqrt {Dt} }}{{\sqrt \pi  }} ,   & t \ll t_d,  \\
\sqrt {\pi Dt} + \frac{{\sqrt \pi  x_0^2}}{{4\sqrt {Dt} }}, &   t \gg t_d.  \\ 
\end{array}  \right. 
\end{eqnarray}
In the short-time limit, $t \ll t_d=\frac{x_0^2}{D}$, the result coincides with the counterpart of the free Brownian motion. In the long-time limit, $t \gg t_d$, the expected maximum is weakly related to the initial position $x_0$, and grows with $t^{1/2}$ as $t$ increases. Interestingly, the expected maximum under the case is  $\pi/2$ times that of the free Brownian motion starting from the origin. 

\begin{figure}
	\centerline{\includegraphics*[width=0.6\columnwidth]{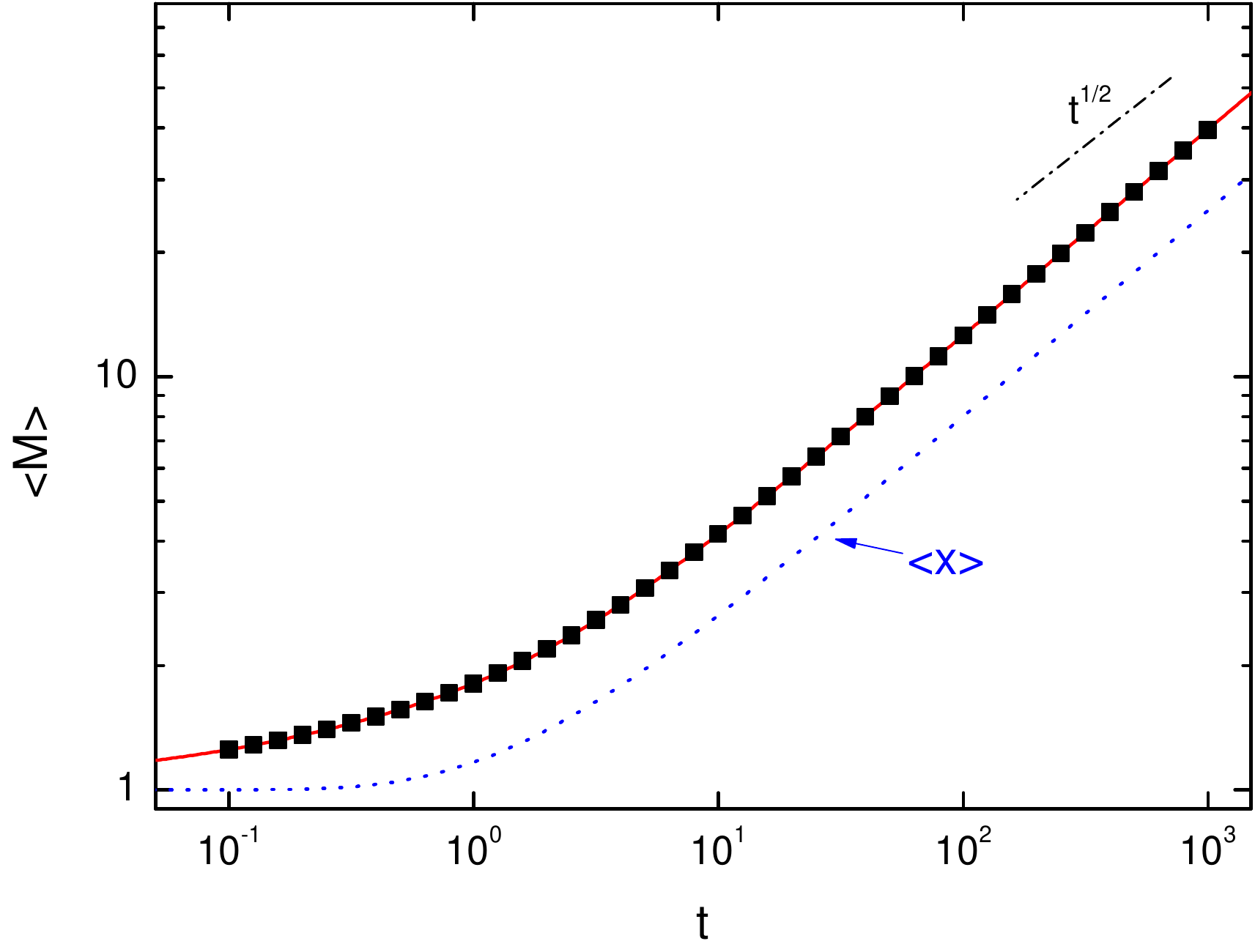}}
	\caption{Expected value $\langle  M(x_0,t)  \rangle$ of the maximum displacement $M$  as a function of $t$, where $x_0=1$ and $D=1/2$ are fixed. Solid line and symbols represent the theory and simulation results, respectively. For comparison, we also plot the expectation $\langle  x(t)  \rangle$ of the displacement $x(t)$, as indicated by the dotted line. \label{fig3}}
\end{figure}

To invert Eq.(\ref{eq1.6}) for the general case, we intend to rewrite Eq.(\ref{eq1.6}) as series expansions. To the end, let us denote by $y=e^{\alpha x_0}$, and use the equations $\arctan ( {1/y} ) = \sum_{n=0}^{\infty}{\frac{{{{\left( { - 1} \right)}^n}}}{{2n + 1}}{y^{ - \left( {2n + 1} \right)}}} $ ($ y\geq 1$) and $\cosh(\ln y)=\frac{1+y^2}{2y}$ to obtain 
\begin{eqnarray}\label{eq1.11}
 \int_0^\infty  dt {\langle {M( x_0, t )} \rangle {e^{ - st}}} &=& \frac{{{x_0}}}{s}   +  \frac{1}{{\alpha s}}\sum\limits_{n = 0}^\infty  {\frac{{{{\left( { - 1} \right)}^n}}}{{2n + 1}}} \left[ {{e^{ - 2n\alpha {x_0}}} + {e^{ - (2n + 2)\alpha {x_0}}}} \right] \nonumber \\ &=&\frac{{{x_0}}}{s} + \frac{1}{{\alpha s}} - \frac{2}{{\alpha s}}\sum\limits_{n = 1}^\infty  {\frac{{{{\left( { - 1} \right)}^{n }}}}{{4{n^2} - 1}}} {e^{ - 2n\alpha {x_0}}}.
\end{eqnarray}
Performing the Laplace transform inversion for Eq.(\ref{eq1.11}), we obtain
\begin{eqnarray}\label{eq1.12}
\langle {M( {{x_0},t} )} \rangle  ={x_0} + 2\sqrt {\frac{{Dt}}{\pi }}  - 4\sum\limits_{n = 1}^\infty  {\frac{{{{\left( { - 1} \right)}^n}}}{{4{n^2} - 1}}} \left[ {\sqrt {\frac{{Dt}}{\pi }} {e^{ - \frac{{{n^2}x_0^2}}{{Dt}}}} - n{x_0}{\rm{erfc}}\left( {\frac{{n{x_0}}}{{\sqrt {Dt} }}} \right)} \right],
\end{eqnarray}
where ${\rm{erfc}}(x)$ is the complementary error function. We note that the result in Eq.(\ref{eq1.12}) was obtained in a previous work as well (see Eq.(5) in \cite{chupeau2015convex}). However, in the present work we additionally obtain the distribution of $M$ and will further derive the statistics of extremal time in the subsequent sections. In Fig.\ref{fig3}, we show a log–log plot of $\langle  M(x_0,t)  \rangle$ as a function of $t$, where we have fixed $x_0=1$ and $D=1/2$. The theoretical prediction is in well agreement with simulations.  For comparison, we plot in Fig.\ref{fig3} the expectation $\langle  x(t)  \rangle$ of the displacement $x(t)$ as a function of $t$ as well. The details of derivation of $\langle  x(t)  \rangle$ can be found in appendix \ref{appen1}. In the short-time limit, $\langle  x(t)  \rangle =x_0$ keeps unchanged as a free Brownian motion, and in the long-time limit $\langle  x(t)  \rangle \sim \sqrt{t}$, but the prefactor is smaller than that of the expected maximum displacement.

\section{Joint distribution $P(M,t_m|x_0,t)$}\label{sec6}
Let us define $P(M,t_m|x_0,t)$ as the joint probability density function that the displacement $x(\tau)$ reaches its maximum $M$ at time $t_m$ with a duration $t$, providing that the Brownian starts from the position $x_0$ ($>0$). To compute the joint distribution $P(M,t_m|x_0,t)$, we can decompose the trajectory into two parts: a left-hand segment for which $0<\tau<t_m$, and a right-hand segment for which $t_m<\tau<t$, as shown in Fig.\ref{fig1}. The statistical weight of the first segment equals to the propagator $G(M,t_m|x_0)$. However, it turns out that $G(M,t_m|x_0)=0$ which implies that the contribution from this part is zero. To circumvent this problem, we compute $G(M-\epsilon,t_m|x_0)$ and later take the limit $\epsilon \to 0^{+}$ \cite{majumdar2004exact}. The statistical weight of the second segment is given by the survival probability $Q(M-\epsilon,t-t_m|M)$. Due to the Markov property, the joint probability density $P(M,t_m|x_0,t)$ can be written as the product of the statistical weights of two segments \cite{randon2007distribution}, 
\begin{eqnarray}\label{eq3.1}
P\left( {M,{t_m}|{x_0},t} \right) = \mathop {\lim }\limits_{\epsilon \to 0^{+}}  \mathcal{N} G\left( {M - \epsilon,{t_m}|{x_0}} \right)Q\left( {M - \epsilon,t - {t_m}|M} \right).
\end{eqnarray}
Here, the normalization factor $\mathcal{N}$ may depend on $D$ and $\epsilon$. In principle, $\mathcal{N}$ depends also on $x_0$ and $t$. However, here we assume and verify a posteriori that it is indeed independent of $x_0$ and $t$.

Let us integrate $P(M,t_m|x_0,t)$ over $t_m$ from 0 to $t$ and over $M$ from $x_0$ to $\infty$, and then perform the Laplace transform with respect to $t$, such that the convolution structure of Eq.(\ref{eq3.1}) can be exploited, 
\begin{eqnarray}\label{eq3.2}
\frac{1}{s}=\int_{0}^{\infty} dt {{e^{ - st}} \int_{{x_0}}^\infty dM  {\int_0^t dt_m P( {M,{t_m}|{x_0},t} )} }  = \mathop {\lim }\limits_{\epsilon \to 0^{+}} \mathcal{N} \int_{x_0}^{\infty}dM\tilde G( {M - \epsilon,s|{x_0}} )\tilde Q( {M - \epsilon,s|M} ) .
\end{eqnarray}
Here, $\tilde G( {M - \epsilon,s|{x_0}} )$ and $\tilde Q( {M - \epsilon,s|M} )$ can be obtained from Eq.(\ref{eq2.8}) and Eq.(\ref{eq1.3}). Up to the leading order of $\epsilon$, they are    
\begin{eqnarray}\label{eq3.3}
\tilde G( {M - \epsilon,{s}|{x_0}} ) = \frac{{\cosh \left( {\alpha {x_0}} \right)}}{{D\cosh \left( {\alpha M} \right)}}\epsilon,
\end{eqnarray}
and
\begin{eqnarray}\label{eq3.4}
\tilde Q\left( {M - \epsilon,s|M} \right) = \frac{{\alpha \tanh \left( {\alpha M} \right)}}{s}\epsilon,
\end{eqnarray}
where $\alpha=\sqrt{s/D}$ again. By inserting Eq.(\ref{eq3.3}) and Eq.(\ref{eq3.4}) into Eq.(\ref{eq3.2}) and completing the integral over $M$ on the right-hand side of Eq.(\ref{eq3.2}), we obtain the  normalization factor $\mathcal{N}$,  
\begin{eqnarray}\label{eq3.5}
\mathcal{N}=\frac{D}{\epsilon^2}.
\end{eqnarray}

Moreover, we perform double Laplace transformations for $P(M,t_m|x_0,t)$ with respect to $t_m$ ($\to s$) and $t$ ($\to p$), and use Eq.(\ref{eq3.3}), Eq.(\ref{eq3.4}) and Eq.(\ref{eq3.5}), to obtain
\begin{eqnarray}\label{eq5.1}
\int_0^\infty  {dt} {e^{ - pt}}\int_0^t {d{t_m}{e^{ - s{t_m}}}P\left( {M,{t_m}|{x_0},t} \right)}  = \frac{{\cosh \left( {\sqrt {\left( {p + s} \right)/D} {x_0}} \right)}}{{\cosh \left( {\sqrt {\left( {p + s} \right)/D} M} \right)}}\frac{{\tanh \left( {\sqrt {p/D} M} \right)}}{{\sqrt {pD} }},
\end{eqnarray}
where we have exchanged the order of integration over $t$ and $t_m$. By using the inverse Laplace transform relations \cite{oberhettinger2012tables},
\begin{eqnarray}\label{eq5.2}
\mathcal{L}_{s \to t}^{ - 1}\left[ {\cosh \left( {\nu {s^{ - 1/2}}} \right) {\rm{sech}} \left( {a{s^{ - 1/2}}} \right)} \right] &=& \frac{1}{a}\frac{\partial }{{\partial \nu }}{\theta _1}\left( {\frac{1}{2}\nu {a^{ - 1}}|t{a^{ - 2}}} \right),\\
\mathcal{L}_{s \to t}^{ - 1}\left[ {{s^{ - 1/2}}\tanh \left( {a{s^{ - 1/2}}} \right)} \right]& =& \frac{1}{a}{\theta _2}\left( {0|t{a^{ - 2}}} \right),
\end{eqnarray}
where
\begin{eqnarray}\label{eq5.3}
{\theta _1}\left( {z|t} \right) & =& 2\sum\limits_{n = 0}^\infty  {{{\left( { - 1} \right)}^n} e^{ - {{\left( {n + 1/2} \right)}^2}{\pi ^2}t}  } \sin \left[ {\left( {2n + 1} \right)\pi z} \right] , \\
{\theta _2}\left( {z|t} \right)& =& 2\sum\limits_{n = 0}^\infty  { e^{ - {{\left( {n + 1/2} \right)}^2}{\pi ^2}t} } \cos \left[ {\left( {2n + 1} \right)\pi z} \right] ,
\end{eqnarray}
are the Elliptic theta functions, we obtain $P(M,t_m|x_0,t)$ from Eq.(\ref{eq5.1}), 
\begin{eqnarray}\label{eq5.4}
P\left( {M,{t_m}|{x_0},t} \right)& =& \frac{1}{M^2}\frac{\partial }{{\partial \nu }}{\theta _1}\left( {\frac{1}{2}\nu {a^{ - 1}}|{t_m}{a^{ - 2}}} \right){\theta_2}\left( {0|(t-t_m){a^{-2}}} \right) \nonumber \\ &=&
\frac{{4\pi D}}{{{M^3}}}\sum\limits_{{n_1} = 0}^\infty  {\sum\limits_{{n_2} = 0}^\infty  {{{\left( { - 1} \right)}^{{n_1}}}\left( {{n_1} + \frac{1}{2}} \right)} {e^{ - \frac{{{\pi ^2}D}}{{{M^2}}}\left[ {{{\left( {{n_1} + \frac{1}{2}} \right)}^2}{t_m} + {{\left( {{n_2} + \frac{1}{2}} \right)}^2}\left( {t - {t_m}} \right)} \right]}}} \cos \left[ {\frac{{\left( {{n_1} + 1/2} \right)\pi {x_0}}}{M}} \right]
\end{eqnarray}
where $a=M/\sqrt{D}$ and $\nu=x_0/\sqrt{D}$. In fact, the joint distribution $P(M,t_m|x_0,t)$ in Eq.(\ref{eq5.4}) can be also obtained directedly from the results in the time domain. This can be done more easily by series expansions of $G(M-\epsilon,t_m|x_0)$ and $Q(M-\epsilon,t-t_m|M)$ to the first order in terms of  Eq.(\ref{eq4.1}) and Eq.(\ref{eq4.2}), and then inserting them into Eq.(\ref{eq3.1}).

\section{Marginal distribution $P(t_m|x_0,t)$}\label{sec7}
Integrating $P(M,t_m|x_0,t)$ in Eq.(\ref{eq5.4}) over $M$ from $x_0$ to $\infty$, we get to the marginal distribution $P(t_m|x_0,t)$ of $t_m$ at which the maximum displacement $M$ is achieved, 
\begin{eqnarray}\label{eq6.1}
P\left( {{t_m}|{x_0},t} \right)& = &\int_{{x_0}}^\infty  {dMP\left( {M,{t_m}|{x_0},t} \right)} \nonumber \\& =& 4\pi D\sum\limits_{{n_1} = 0}^\infty  {\sum\limits_{{n_2} = 0}^\infty  {{{\left( { - 1} \right)}^{{n_1}}}\left( {{n_1} + \frac{1}{2}} \right)} } F\left( { {\pi ^2}D\left[ {{{\left( {{n_1} + \frac{1}{2}} \right)}^2}{t_m} + {{\left( {{n_2} + \frac{1}{2}} \right)}^2}\left( {t - {t_m}} \right)} \right], \left( {{n_1} + \frac{1}{2}} \right)\pi {x_0}} \right),
\end{eqnarray}
where 
\begin{eqnarray}\label{eq6.2}
F\left( {a,b} \right) &=& \int_{{x_0}}^\infty  dM\frac{1}{{{M^3}}}{e^{ - \frac{a}{{{M^2}}}}}\cos \left( {\frac{b}{M}} \right) \nonumber \\ &=& \frac{1}{{2a}} - \frac{1}{{2a}}{e^{ - \frac{a}{{x_0^2}}}}\cos \left( {\frac{b}{{{x_0}}}} \right) - \frac{{\sqrt \pi  b}}{{4{a^{3/2}}}}{e^{ - \frac{{{b^2}}}{{4a}}}}{\rm{erfi}}\left( {\frac{b}{{2\sqrt a }}} \right) + \frac{{\sqrt \pi  b}}{{8{a^{3/2}}}}{e^{ - \frac{{{b^2}}}{{4a}}}}\left[ {{\rm{erfi}}\left( {\frac{{2a {\rm{i}}  + b{x_0}}}{{2\sqrt a {x_0}}}} \right) - {\rm{erfi}}\left( {\frac{{2a {\rm{i}} - b{x_0}}}{{2\sqrt a {x_0}}}} \right)} \right].
\end{eqnarray}
Here ${\rm{erfi}}(z)=-{\rm{i}}{\rm{erf}}({\rm{i}}z)$ is the imaginary error function, and ${\rm{i}}$ is the unit imaginary number.

\begin{figure}
	\centerline{\includegraphics*[width=0.8\columnwidth]{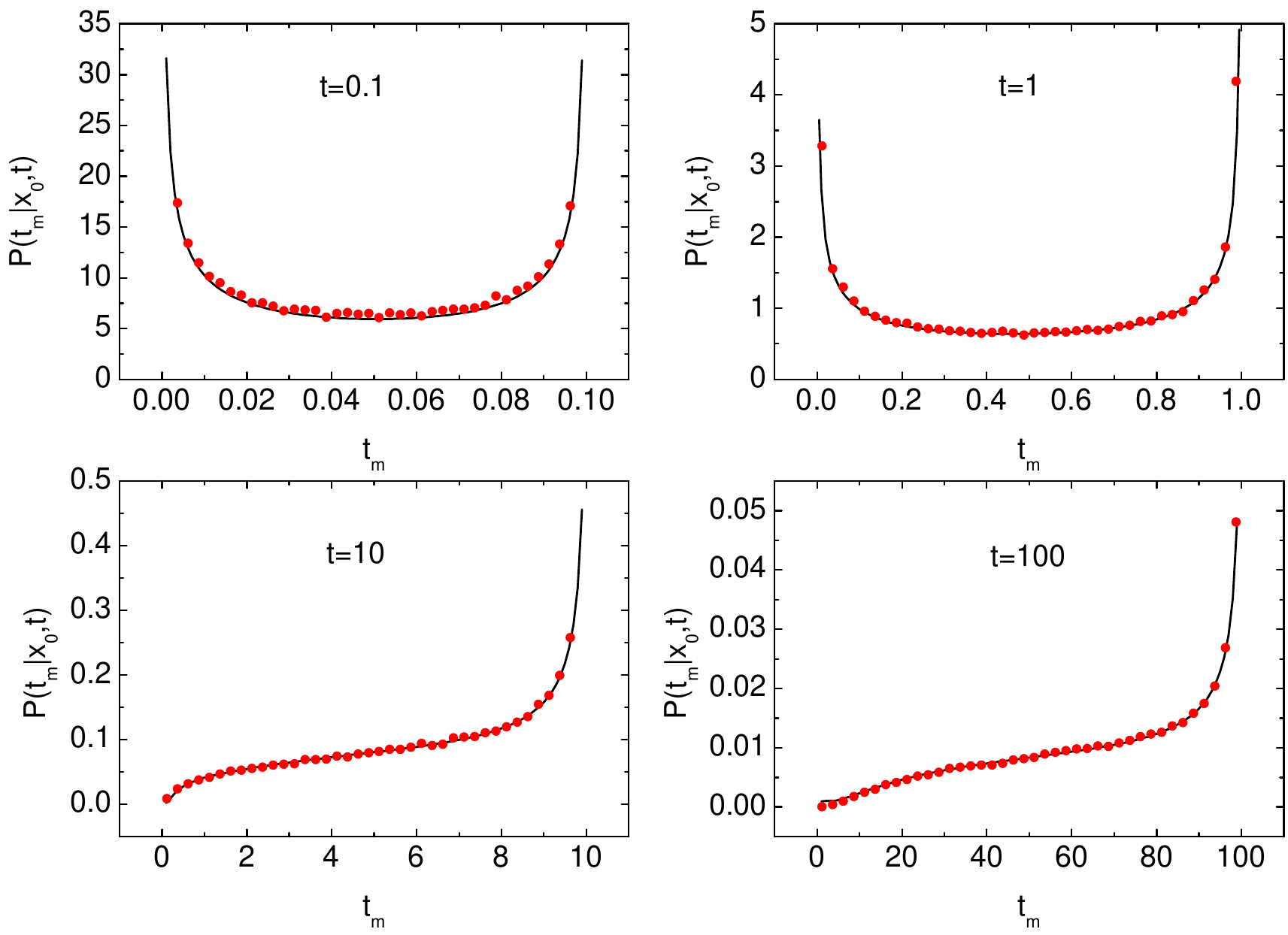}}
	\caption{The distribution $P(t_m|x_0,t)$ of $t_m$ at which the displacement reaches its maximum for different values of $t$, where $x_0=1$ and $D=1/2$ are fixed.  Lines and symbols represent the theory and simulation results, respectively. \label{fig5}}
\end{figure}

In Fig.\ref{fig5}, we show the distribution $P(t_m|x_0,t)$ of $t_m$ for several different values of $t$. The results from theory and simulations are in agreement well.  For $t \ll t_d$, the particle has a very low probability of diffusing the reflective boundary, such that the particle behaves a free diffusion without any boundaries and $P(t_m|x_0,t)$ is U-shaped as indicated in Eq.(\ref{eq0.3}). For $t \gg t_d$, the reflective boundary comes into effect and $P(t_m|x_0,t)$ deviates from the U-shaped distribution, and becomes asymmetric with respect to $t_m=t/2$. There is a higher probability of $t_m$ that occurs near $t$.

Moreover, we seek to find the marginal distribution $P(t_m|x_0,t)$ in the Laplace domain. To the end, we perform double Laplace transforms for $P(t_m|x_0,t)$, and use Eq.(\ref{eq3.3}), Eq.(\ref{eq3.4}) and Eq.(\ref{eq3.5}) to obtain
\begin{eqnarray}\label{eq6.3}
\int_0^\infty  {dt} {e^{ - pt}}\int_0^t {d{t_m}{e^{ - s{t_m}}}P\left( {{t_m}|{x_0},t} \right)}  = \int_{{x_0}}^\infty  {dM\frac{{\cosh \left( {\sqrt {\left( {p + s} \right)/D} {x_0}} \right)}}{{\cosh \left( {\sqrt {\left( {p + s} \right)/D} M} \right)}}\frac{{\tanh \left( {\sqrt {p/D} M} \right)}}{{\sqrt {pD} }}} .
\end{eqnarray}
In the short-time limit ($p \to \infty$), using $\cosh(x) \sim e^{x}/2$ and $\tanh(x) \sim 1$ ($x \to \infty$), one easily recovers to the result in Eq.(\ref{eq0.3}) from Eq.(\ref{eq6.3}). Furthermore, one observes that Eq.(\ref{eq6.3}) is not invariant under the exchange of $p+s \leftrightarrow p$. This implies that $P(t_m|x_0,t)$ is no longer symmetric with respect to $t/2$ in the presence of the reflective boundary.  

Taking the derivative of Eq.(\ref{eq6.3}) with respect to $s$ and letting $s \to 0$, we obtain the Laplace transformation of the expectation of $t_m$,
\begin{eqnarray}\label{eq3.8}
\int_0^\infty  {dt{e^{ - pt}}} \langle {{t_m}(x_0,t)} \rangle &=&  - \int_{{x_0}}^\infty  {dM\frac{\partial }{{\partial s}}{{\left[ {\frac{{\cosh \left( {\sqrt {\left( {p + s} \right)/D} {x_0}} \right)}}{{\cosh \left( {\sqrt {\left( {p + s} \right)/D} M} \right)}}} \right]}_{s \to 0}}\frac{{\tanh \left( {\sqrt {p/D} M} \right)}}{{\sqrt {pD} }}} \nonumber \\ & =& \frac{1}{{2pD}}\int_{{x_0}}^\infty  {dM\frac{{\sinh \left( {\sqrt {p/D} M} \right)}}{{{{\cosh }^2}\left( {\sqrt {p/D} M} \right)}}\left[ {M\cosh \left( {\sqrt {p/D} {x_0}} \right)\tanh \left( {\sqrt {p/D} M} \right) - {x_0}\sinh \left( {\sqrt {p/D} {x_0}} \right)} \right]}  \nonumber \\  &=& \frac{1}{{4{p^2}}} - \frac{{{x_0}\cosh \left( \sqrt{p/D}x_0  \right) }}{{4\sqrt {{p^3}D} }} {\rm{arg}}\left( {\frac{{{e^{\sqrt{p/D}x_0}} - {\rm{i}} }}{{{e^{\sqrt{p/D}x_0}} + {\rm{i}} }}} \right) +  \frac{{\cosh \left( \sqrt{p/D}x_0  \right) {e^{ - \sqrt{p/D}x_0}}}}{{8{p^2}}} \nonumber \\ && \times \Phi \left( { - {e^{ - 2\sqrt{p/D}x_0}},2,\frac{1}{2}} \right)  - \frac{{{x_0}}}{{4\sqrt {{p^3}D} }}\tanh \left( \sqrt{p/D}x_0  \right) ,
\end{eqnarray}
where ${\rm{arg}}(z)$ denotes the argument of a complex number $z$, ${\rm{i}}$ is the unit imaginary number, and
\begin{eqnarray}\label{eq3.9.1}
\Phi \left( {z,2,\frac{1}{2}} \right) = \sum\limits_{k = 0}^\infty  {\frac{{{z^k}}}{{{{\left( {k + 1/2} \right)}^2}}}} ,
\end{eqnarray}
is the Lerch transcendent function. It turns out that inverting Eq.(\ref{eq3.8}) with respect to $p$ is a highly nontrivial task. However, in the short-time limit ($p \to \infty$) and in the long-time limit ($p \to 0$), Eq.(\ref{eq3.8}) can be simplified to 
\begin{eqnarray}\label{eq3.10}
\int_0^\infty  {dt{e^{ - pt}}} \langle {{t_m}} \rangle  = \left\{ \begin{array}{cc}
\frac{1}{2p^2} ,   & p \to \infty,  \\
\frac{{1 + 2G}}{{4{p^2}}}, &   p \to 0,  \\ 
\end{array}  \right. 
\end{eqnarray}
where $G=\sum_{k=0}^{\infty}(-1)^k/(2k+1)^2 \approx 0.916$ is the Catalan's constant. Inverting Eq.(\ref{eq3.10}) with respect to $p$, we obtain 
\begin{eqnarray}\label{eq3.11}
\frac{{\langle {{t_m}} \rangle }}{t}  = \left\{ \begin{array}{cc}
\frac{1}{2} ,   & t \ll t_d,  \\
 \frac{{1 + 2G}}{4}, &   t \gg t_d.  \\ 
\end{array}  \right. 
\end{eqnarray}

\begin{figure}
	\centerline{\includegraphics*[width=0.6\columnwidth]{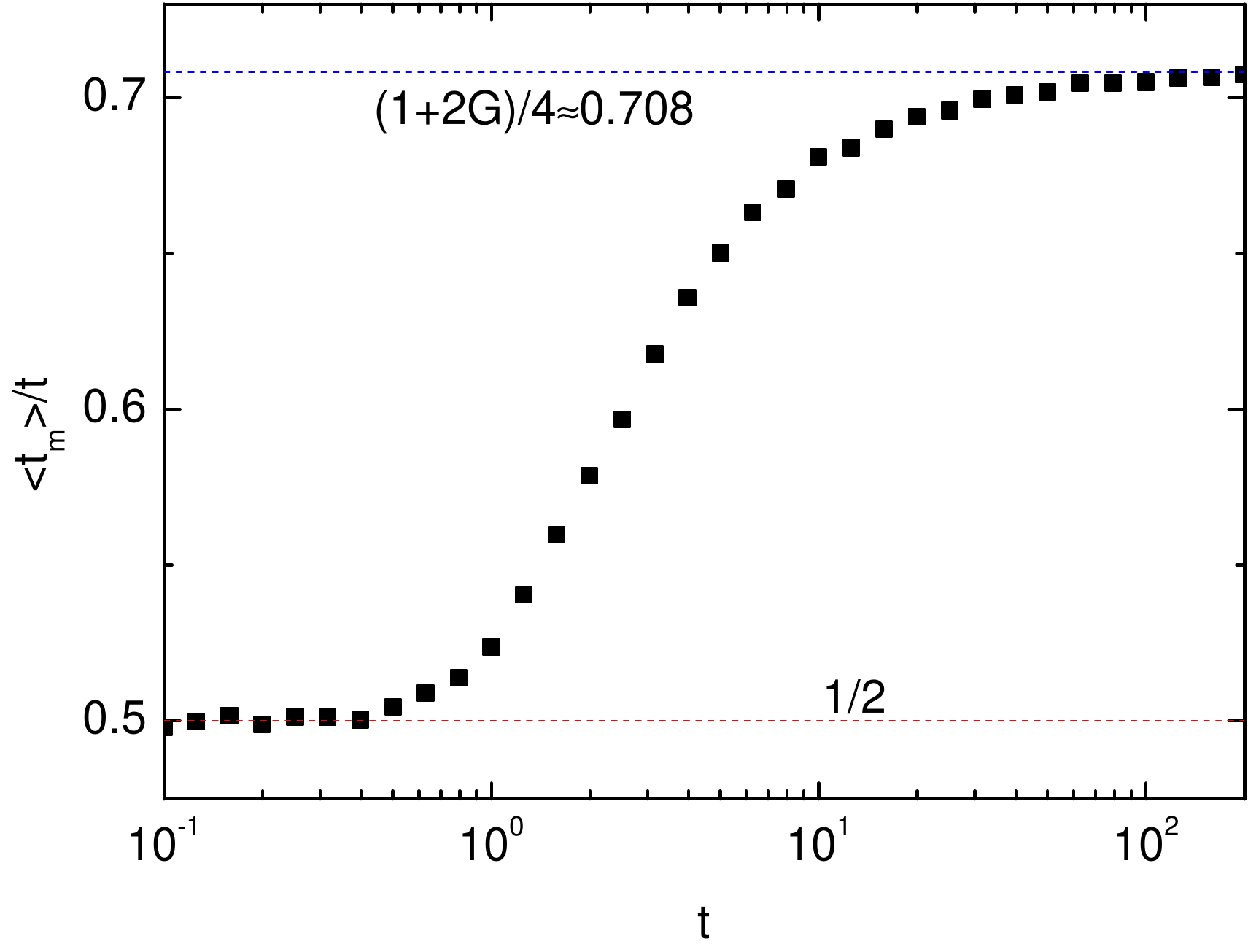}}
	\caption{${\langle {t_m(x_0,t)} \rangle }/t$  as a function of $t$, where $x_0=1$ and $D=1/2$ are fixed. In the short-time limit, $t\ll t_d=\frac{x_0^2}{D}$, ${\langle {t_m} \rangle }/t \to \frac{1}{2}$, and in the long-time limit, $t\gg t_d$, ${\langle {t_m} \rangle }/t \to \frac{{1 + 2G}}{4}\approx 0.708$, where $G\approx 0.916$ is the Catalan's constant. \label{fig4}}
\end{figure}

Coincidentally, we realize that the Catalan's constant $G$  appears also in the expression of the probability of fractional Brownian motion to first exit at the upper boundary in an interval \cite{PhysRevE.99.032106}. However, there is no a direct connection between them at present.   In Fig.\ref{fig4}, we plot $\langle{t_m}\rangle/t$ as a function of $t$ for the fixed $x_0=1$ and $D=1/2$, from which we can see that the ratio $\langle{t_m}\rangle/t$ increases monotonically with $t$. $\langle{t_m}\rangle/t$ approaches $1/2$ in the short-time limit, and tends towards $\frac{{1 + 2G}}{4}\approx 0.708$ in the long-time limit, which are consistent with our theoretical predictions in Eq.(\ref{eq3.11}).

\section{The statistics of the last time $t_\ell$ the process visits the starting position $x_0$ and the occupation time $t_o$ spent above $x_0$}\label{sec8}
As mentioned in the Introduction, for a free Brownian motion the last time $t_\ell$ that the process visits the starting position and the occupation time $t_o$ spent above the starting position over a duration $t$ are also distributed by Eq.(\ref{eq0.3}), called the other two arcsine laws. In this section, we will show how the two arcsine laws are modified in the presence of a reflective wall at the origin.

\subsection{The statistics of $t_\ell$}
Using the path decomposition technique as done in Sec.\ref{sec5}, we can obtain the distribution of the last time $t_\ell$ when a stochastic path visits the starting position $x_0$. We decompose the path into two parts: (i) the process starts from $x_0$($>0$) at $t=0$ and returns 
to the starting position at $t_\ell$. The probability density is given by the propagator $G_0(x_0,t_\ell|x_0)$ (here the propagator will be derived only in the presence of a reflective boundary, different from Eq.(\ref{eq4.1}) where a reflective boundary and an absorbing boundary are both present); (ii) In the time interval from $t_\ell$ to $t$, the process never returns to the starting position. The probability is given by survival probability $Q(x_0,t-t_\ell|x_0)$. Since the process is Markovian, the two parts are statistically independent. Since $Q(x_0,t-t_\ell|x_0)=0$ and thus the probability of the second part is equal to zero. To overcome this difficulty, we consider the probability as $Q(x_0+\epsilon, t-t_\ell|x_0) + Q(x_0-\epsilon, t-t_\ell|x_0)$ and then take the $\epsilon \to 0^{+}$ limit. Thus, the probability density function of $t_\ell$ can be expressed as
\begin{eqnarray}\label{eq9.1}
P(t_\ell|t)=\mathop {\lim }\limits_{\epsilon \to 0^{+}}  \mathcal{N} G_0(x_0,t_\ell|x_0) \left[ Q(x_0+\epsilon, t-t_\ell|x_0) + Q(x_0-\epsilon, t-t_\ell|x_0) \right] .
\end{eqnarray}
The propagator $G_0(x,t|x_0)$ in the presence of a reflective wall at $x=0$ can be easily obtain from the method of image, given by
\begin{eqnarray}\label{eq9.2}
G_0(x,t|x_0)=\frac{1}{\sqrt{4 \pi D t}} \left[ e^{-\frac{(x-x_0)^2}{4 D t}} +e^{-\frac{(x+x_0)^2}{4 D t}} \right]  .
\end{eqnarray}
The survival probability $Q(x_0,t|M)$ should be considered separately for the case $M<x_0$ and for the case $M>x_0$. When the absorbing boundary $M$ is below the starting position $x_0$, $M<x_0$, the reflective boundary at the origin does not produce any  effect, and thus the survival probability $Q(x_0,t|M)$ is a well-known result \cite{redner2001guide}, 
\begin{eqnarray}\label{eq9.3}
Q(x_0,t|M<x_0)={\rm{erf}}\left( \frac{x_0-M}{2\sqrt{Dt}} \right)   .
\end{eqnarray}
For $M>x_0$, the role of the reflective wall at $x=0$ needs to take into account. In fact, under this case the survival probability $Q(x_0,t|M>x_0)$ has been obtained in Sec.\ref{sec4}, see Eq.(\ref{eq4.2}).

It is useful to perform the double Laplace transformations for $P(t_\ell|t)$ with respect to $t_\ell$ ($\to s$) and with respect to $t$ ($\to p$), 
\begin{eqnarray}\label{eq9.4}
\int_{0}^{\infty}dt e^{-p t} \int_{0}^{t}d t_\ell e^{-s t_\ell} P(t_\ell|t)=\mathop {\lim }\limits_{\epsilon \to 0^{+}}  \mathcal{N} \tilde{G}_0(x_0, p+s |x_0) \left[ \tilde{Q}(x_0+\epsilon, p|x_0) + \tilde{Q}(x_0-\epsilon, p|x_0) \right] ,
\end{eqnarray}
where we have exchanged the orders of the integrals over $t$ and $t_\ell$, and made the  variable substitution $t'=t-t_\ell$. In principle, $\mathcal{N}$ can also depend on $t$. However, here we assume and verify a posteriori that it is indeed independent of $t$. $\tilde{G}_0(x_0, p |x_0)$ and $\tilde{Q}(x_0+\epsilon, p|x_0)$ can be obtained from Eq.(\ref{eq9.2}) and Eq.(\ref{eq9.3}), given by 
\begin{eqnarray}\label{eq9.5}
\tilde{G}(x_0, p |x_0)=\frac{{1 + {e^{ - 2\sqrt {p/D} {x_0}}}}}{{2\sqrt {Dp} }},
\end{eqnarray} 
and 
\begin{eqnarray}\label{eq9.6}
\tilde{Q}(x_0+\epsilon, p|x_0)=\frac{\epsilon}{\sqrt{Dp}}+o(\epsilon^2).
\end{eqnarray} 
$\tilde{Q}(x_0-\epsilon, p|x_0)$ can be obtained from Eq.(\ref{eq1.3}),
\begin{eqnarray}\label{eq9.7}
\tilde{Q}(x_0-\epsilon, p|x_0)=\frac{{\tanh \left( {\sqrt {p/D} {x_0}} \right)}}{{\sqrt {Dp} }}\epsilon+o(\epsilon^2)
\end{eqnarray} 
Letting $s \to 0$, the left-hand side of Eq.(\ref{eq9.4}) is just equal to $1/p$ due to the normalization condition, $\int_{0}^{t} d t_\ell P(t_\ell|t)=1$. We substitute Eq.(\ref{eq9.5}), Eq.(\ref{eq9.6}) and Eq.(\ref{eq9.7}) into the right-hand side of Eq.(\ref{eq9.4}), and then compare both sides of Eq.(\ref{eq9.4}), which yields
\begin{eqnarray}\label{eq9.8}
\mathcal{N}=\frac{D}{\epsilon}.
\end{eqnarray}

In the time domain, we have
\begin{eqnarray}\label{eq9.9}
G(x_0,t_\ell|x_0)=\frac{{1 + {e^{ - \frac{{x_0^2}}{{D{t_\ell}}}}}}}{{\sqrt {4\pi D{t_\ell}} }},
\end{eqnarray}
\begin{eqnarray}\label{eq9.10}
Q(x_0+\epsilon, t-t_\ell|x_0)={\frac{\epsilon}{{\sqrt {\pi D\left( {t - {t_\ell}} \right)} }}} +o(\epsilon^2),
\end{eqnarray}
and 
\begin{eqnarray}\label{eq9.11}
Q(x_0-\epsilon, t-t_\ell|x_0)=\frac{\epsilon}{{{x_0}}}{\theta _2}\left( {0\left| {\frac{{D\left( {t - {t_\ell}} \right)}}{{x_0^2}}} \right.} \right) +o(\epsilon^2)  = 
\frac{2\epsilon}{{{x_0}}}\sum\limits_{n = 0}^\infty  {{e^{ - \frac{{{{\left( {n + 1/2} \right)}^2}{\pi ^2}D\left( {t - {t_\ell}} \right)}}{{x_0^2}}}}}+o(\epsilon^2) ,
\end{eqnarray}
where ${\theta_2}( {z| t })$ is the Elliptic theta function defined in Eq.(\ref{eq5.3}). Therefore, the distribution of $t_\ell$ can be obtained by inserting Eqs.(\ref{eq9.8}), (\ref{eq9.9}), (\ref{eq9.10}) and (\ref{eq9.11}) into Eq.(\ref{eq9.1}), 
\begin{eqnarray}\label{eq9.12}
P(t_\ell|t)=\sqrt {\frac{D}{{4\pi }}} \frac{{1 + {e^{ - \frac{{x_0^2}}{{D{t_\ell}}}}}}}{{\sqrt {{t_\ell}} }}\left[ {\frac{1}{{\sqrt {\pi D\left( {t - {t_\ell}} \right)} }} + \frac{1}{{{x_0}}}{\theta _2}\left( {0\left| {\frac{{D\left( {t - {t_\ell}} \right)}}{{x_0^2}}} \right.} \right)} \right] .
\end{eqnarray}

\begin{figure}
	\centerline{\includegraphics*[width=0.6\columnwidth]{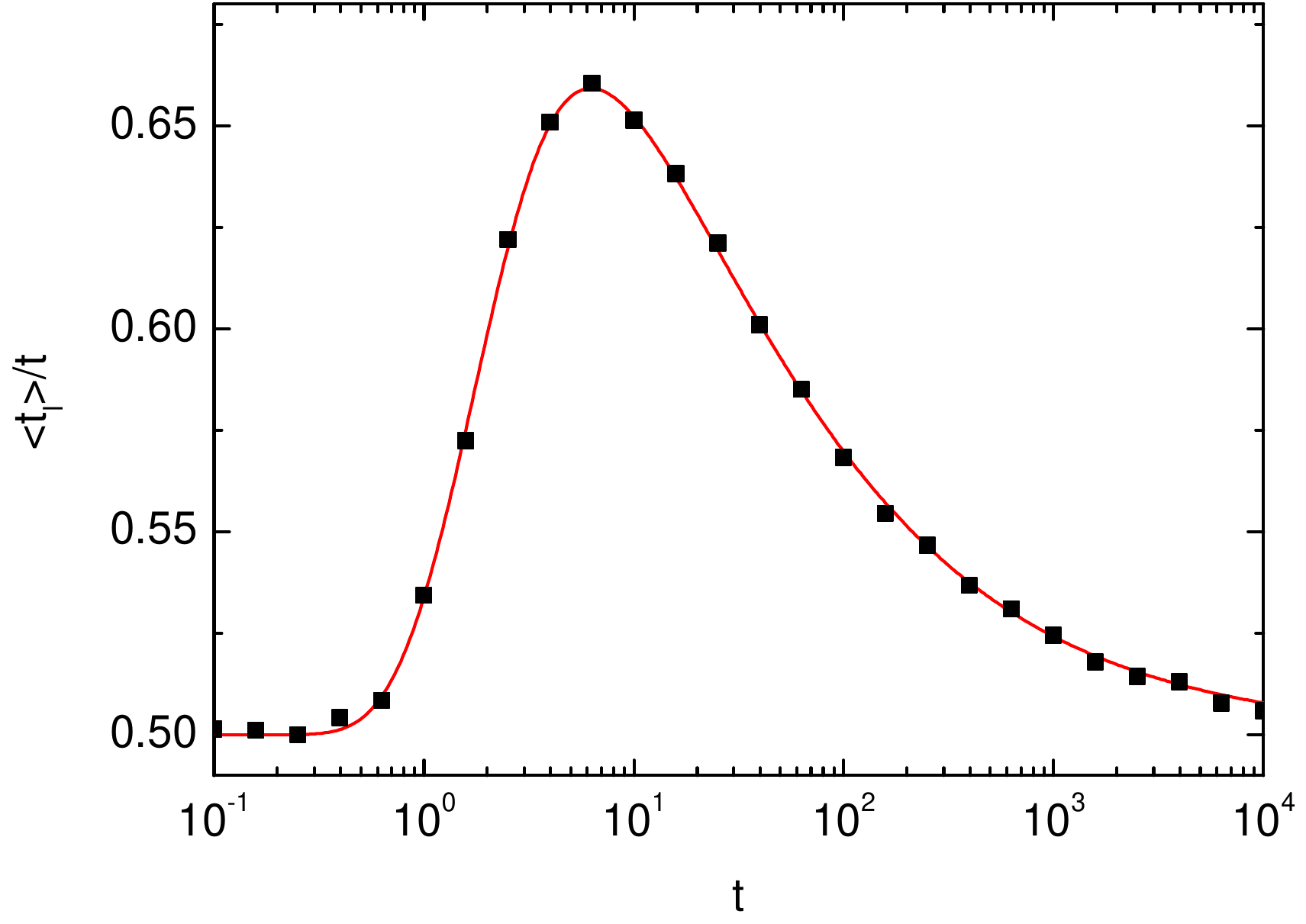}}
	\caption{${\langle {t_\ell(x_0,t)} \rangle }/t$ as a function of $t$, where $x_0=1$ and $D=1/2$ are fixed. Line and symbols correspond to the theoretical and simulation results, respectively.  \label{fig7}}
\end{figure}

By taking the derivative of Eq.(\ref{eq9.4}) with respect to $s$ and then letting $s \to 0$, we obtain the Laplace transform of the expectation of $t_\ell$
\begin{eqnarray}\label{eq9.13}
\int_{0}^{\infty} dt e^{-pt} \langle t_\ell(x_0,t) \rangle &=&\frac{1}{{2{p^2}}} + \frac{{{x_0}}}{{\sqrt {D{p^3}} \left( {1 + {e^{2\sqrt {p/D} {x_0}}}} \right)}} \nonumber \\ &=&
\frac{1}{{2{p^2}}} + \frac{{{x_0}}}{{\sqrt {D{p^3}} }}\sum\limits_{n = 0}^\infty  {{{\left( { - 1} \right)}^n}{e^{2\left( {n + 1} \right)\sqrt {p/D} {x_0}}}} .
\end{eqnarray}
The series representation in the last line of Eq.(\ref{eq9.13}) is to seek conveniently the inverse Laplace transform with respect to $p$. For this purpose, we have
\begin{eqnarray}\label{eq9.14}
{\langle t_\ell(x_0,t) \rangle }=\frac{t}{2} + \frac{{2{x_0}}}{D}\sum\limits_{n = 0}^\infty  {{{\left( { - 1} \right)}^n}} \left[ {\sqrt {\frac{{Dt}}{\pi }} {e^{ - \frac{{{{\left( {n + 1} \right)}^2}x_0^2}}{{Dt}}}} - \left( {n + 1} \right){x_0}{\rm{erfc}}\left( {\frac{{\left( {n + 1} \right){x_0}}}{{\sqrt {Dt} }}} \right)} \right],
\end{eqnarray}
where ${\rm{erfc}}(x)$ is the complementary error function. In Fig.\ref{fig7}, we plot $\langle t_\ell \rangle/t$ as a function of $t$ and compared it with the numerical simulations to find an excellent match. Quite interestingly, $\langle t_\ell \rangle/t$ exhibits a non-monotonic behaviour with respect to $t$. In the limits $t \to 0$ and $t \to \infty$, $\langle t_\ell \rangle/t \to 1/2$. At an intermediate value of $t$, $t \approx 3.1216 x_0^2/D$, $\langle t_\ell \rangle/t$ attains its maximum, $\langle t_\ell \rangle/t \approx 0.659$.

\subsection{The statistics of $t_o$}    
Let us denote by $t_o$ the occupation time the Brownian process $x(\tau)$ spent above the position $x_0$ during time $t$, providing that the process starts from the position $y_0$ in the presence of a reflective boundary at the origin. The occupation time $t_o$ is conveniently measured by the following the Brownian functional, 
\begin{eqnarray}\label{eq10.1}
t_o=\int_{0}^{t} \Theta(x(\tau)-x_0)d \tau,
\end{eqnarray}
where $\Theta(x)$ is the Heaviside step function as defined before. Clearly, $t_o$ is a random variable taking different values for different Brownian paths. Let us define $P(t_o|t)$ as the probability density function of $t_o$. Since $t_o$ has only positive support, a natural step is to introduce the Laplace transform of $P(t_o|t)$, 
\begin{eqnarray}\label{eq10.2}
Q(y_0, s, t)=\int_{0}^{t} e^{-s t_o} P(t_o|t) \equiv \langle e^{-s \int_{0}^{t} \Theta(x(\tau)-x_0)d \tau} \rangle,
\end{eqnarray}
where $\langle \cdot \rangle$ denotes the expectation with respect to the process $x(\tau)$ with started at $y_0$. $Q(y_0, s, t)$ evolves according to the backward equation \cite{majumdar2007brownian},
\begin{eqnarray}\label{eq10.3}
\frac{\partial Q(y_0, s, t)}{\partial t}=D \frac{\partial^2 Q(y_0, s, t)}{\partial y_0^2}-s \Theta(y_0-x_0) Q(y_0, s, t).
\end{eqnarray}
Performing a further Laplace transform for $Q(y_0, s, t)$ with respect to $t$, $\tilde{Q}(y_0,s,p)=\int_{0}^{\infty}dt e^{-pt}Q(y_0, s, t)$, Eq(\ref{eq10.3}) becomes
\begin{eqnarray}\label{eq10.4}
D \frac{d^2 \tilde{Q}(y_0,s,p)}{d y_0^2}-\left[ s \Theta(y_0-x_0)+ p \right] \tilde{Q}(y_0,s,p)=-1.
\end{eqnarray}
Eq.(\ref{eq10.4}) can be solved for the regions $y_0<x_0$ and $y_0>x_0$, separately, which yields, 
\begin{eqnarray}\label{eq10.5}
\tilde{Q}(y_0,s,p) = \left\{ \begin{array}{ll}
	\frac{1}{p} +A_1 e^{\sqrt{p/D}y_0}+B_1 e^{-\sqrt{p/D}y_0},   & y_0 \leq x_0,  \\
	\frac{1}{p+s}+A_2 e^{\sqrt{(p+s)/D}y_0}+B_2 e^{-\sqrt{(p+s)/D}y_0}, & y_0>x_0,  \\ 
	\end{array}  \right. 
\end{eqnarray}
where four unknown coefficients $A_1$, $B_1$, $A_2$ and $B_2$ can be determined as follows. If $y_0 \to \infty$, the particle will spend all its time above the position $x_0$, such that $P(t_o|t)=\delta(t_o-t)$ and thus $\tilde{Q}(y_0,s,p)=1/(p+s)$. It is clear that this boundary condition leads to $A_2=0$. The zero-flux boundary condition at the origin requires $\partial _{y_0} \tilde{Q}(y_0,s,p)|_{y_0=0}=0$, which leads to $A_1=B_1$. The matching conditions at $y_0=x_0$ are $\tilde{Q}(x_0^{+},s,p)=\tilde{Q}(x_0^{-},s,p)$ and $\partial _{y_0} \tilde{Q}(y_0,s,p)|_{y_0=x_0^{+}}=\partial _{y_0} \tilde{Q}(y_0,s,p)|_{y_0=x_0^{-}}$, which yields
\begin{eqnarray}\label{eq10.6}
\left\{ \begin{array}{l}
	{A_1} =  - \frac{s}{{2p\sqrt {p + s} \left[ {\sqrt {p + s} \cosh \left( {\sqrt {p/D} {x_0}} \right) + \sqrt p \sinh \left( {\sqrt {p/D} {x_0}} \right)} \right]}}\\
	{B_2} = \frac{{s\sinh \left( {\sqrt {p/D} {x_0}} \right)\left[ {\sinh \left( {\sqrt {\left( {p + s} \right)/D} {x_0}} \right) + \cosh \left( {\sqrt {\left( {p + s} \right)/D} {x_0}} \right)} \right]}}{{\sqrt p \left( {p + s} \right)\left[ {\sqrt {p + s} \cosh \left( {\sqrt {p/D} {x_0}} \right) + \sqrt p \sinh \left( {\sqrt {p/D} {x_0}} \right)} \right]}}
\end{array} \right.
\end{eqnarray}

\begin{figure}
	\centerline{\includegraphics*[width=0.6\columnwidth]{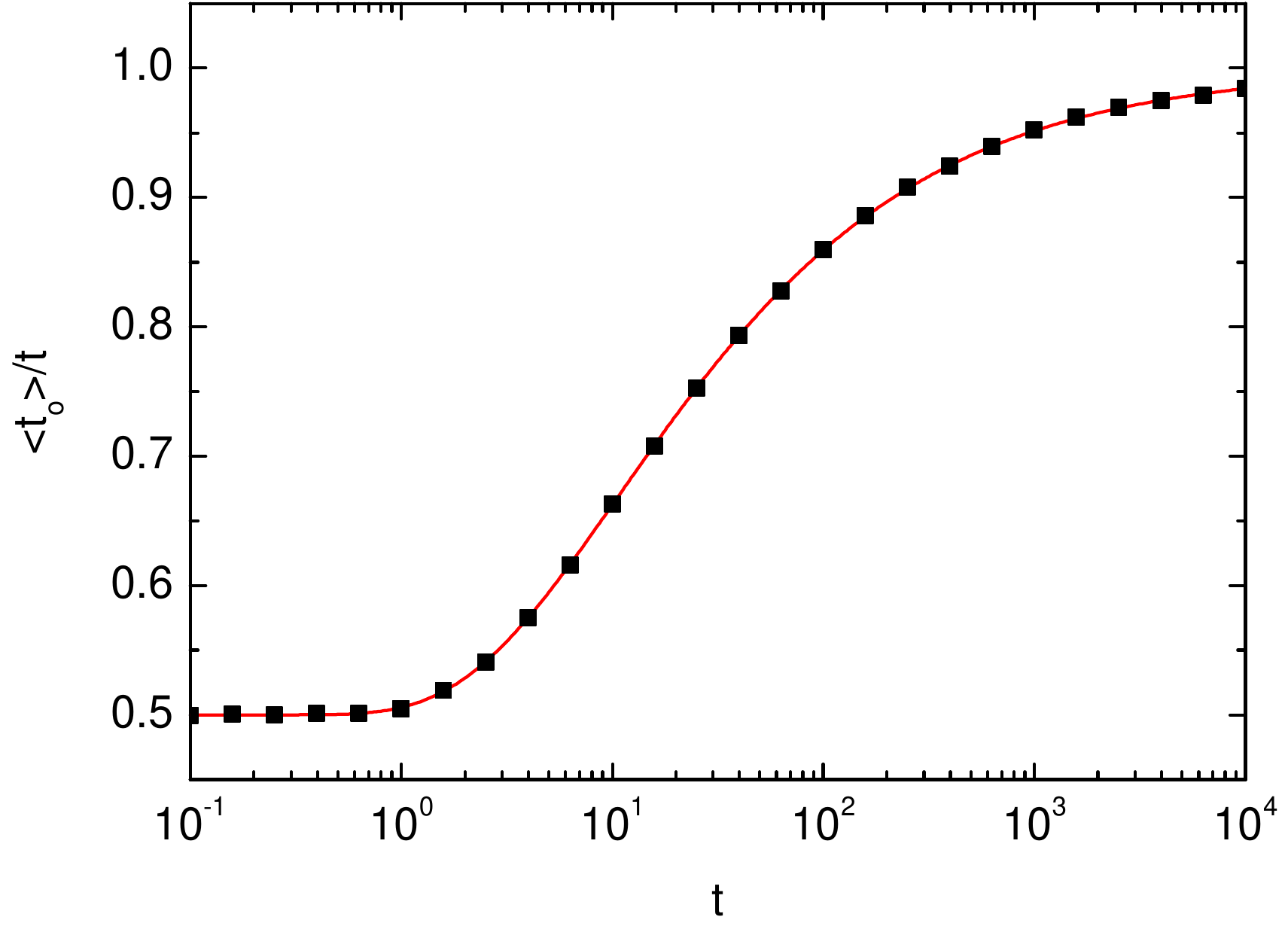}}
	\caption{The fraction of occupation time spent above the starting position $x_0$, ${\langle {t_o(x_0,t)} \rangle }/t$, as a function of $t$, where $x_0=1$ and $D=1/2$ are fixed. Line and symbols correspond to the theoretical and simulation results, respectively. \label{fig6}}
\end{figure}

For the case we are interested in, $y_0=x_0$, we have
\begin{eqnarray}\label{eq10.7}
\tilde{Q}(x_0,s,p)=
\frac{{\sqrt p \cosh \left( {\sqrt {p/D} {x_0}} \right) + \sqrt {p + s} \sinh \left( {\sqrt {p/D} {x_0}} \right)}}{{\sqrt {p\left( {p + s} \right)} \left[ {\sqrt {p + s} \cosh \left( {\sqrt {p/D} {x_0}} \right) + \sqrt p \sinh \left( {\sqrt {p/D} {x_0}} \right)} \right]}}.
\end{eqnarray}
For $x_0 \to \infty$, i.e., when the starting point is far away from the reflective boundary, Eq.(\ref{eq10.7}) reduces to $\tilde{Q}(x_0\to \infty,s,p)=1/\sqrt{p(p+s)}$, and recovers to the classical arcsine law for the occupation time. 

Taking the derivative of $\tilde{Q}(x_0,s,p)$ with respect to $s$ and then letting $s \to 0$, we obtain the Laplace transform of the mean occupation time, 
\begin{eqnarray}\label{eq10.8}
\int_0^\infty  {dt{e^{ - pt}}\langle {{t_o}\left( {{x_0},t} \right)} \rangle }  = -{\left. {\frac{{\partial \tilde Q\left( {{x_0},s,p} \right)}}{{\partial s}}} \right|_{s \to 0}} = \frac{{\cosh \left( {\sqrt {p/D} {x_0}} \right)}}{{{p^2}\left[ {\cosh \left( {\sqrt {p/D} {x_0}} \right) + \sinh \left( {\sqrt {p/D} {x_0}} \right)} \right]}}.
\end{eqnarray}
The inversion of Eq.(\ref{eq10.8}) to obtain the mean occupation time, 
\begin{eqnarray}\label{eq10.9}
\langle {{t_o}\left( {{x_0},t} \right)} \rangle  = \frac{1}{2}\left[ {t - \sqrt {\frac{{4x_0^2t}}{{\pi D}}} {e^{ - \frac{{x_0^2}}{{Dt}}}} + \left( {t + \frac{{2x_0^2}}{D}} \right){\rm{erfc}}\left( {\frac{{{x_0}}}{{\sqrt {Dt} }}} \right)} \right],
\end{eqnarray}
where ${\rm{erfc}}(x)$ is the complementary error function. In Fig.\ref{fig6}, we show the fraction of occupation time spent above the starting position $x_0$, ${\langle {t_o(x_0,t)} \rangle }/t$, as a function of $t$, where $x_0=1$ and $D=1/2$ are fixed.  The simulation result agrees well with the theoretical formula in Eq.(\ref{eq10.9}). In the short-time limit, $t \ll t_d$, the reflective boundary does not come into effect, and thus ${\langle {t_o} \rangle }/t \to 1/2$ as predicted by the arcsine law. As $t$ increases, ${\langle {t_o} \rangle }/t$ increases monotonically, and tends to $1$ in the long-time limit $t \gg t_d$. This is because that under the influence of the reflective wall the process has a higher probability of staying above the starting position $x_0$ for a longer time $t$. The probability can be quantified by integrating the propagator in Eq.(\ref{eq9.2}) over position from $x_0$ to infinity, $\int_{x_0}^{\infty} dx G(x,t|x_0)=1-\frac{1}{2}{\rm{erf}}(x_0/\sqrt{Dt})$, from which one can see that the probability is an increasing function of $t$.

\section{Conclusions}\label{sec9}
We have shown how the presence of a reflective boundary affects the extremal statistics of a one-dimensional Brownian motion over a fixed duration $t$. We analytical obtain the joint and marginal distributions of the maximum displacement $M$ and the time $t_m$ at which the maximum $M$ is reached. As expected, in the short-time limit ($t \ll t_d$), the reflective wall does not produce much impact on the statistics of $M$ and $t_m$, and thus our model is similar to the free Brownian motion. When $t$ becomes comparable with $t_d$, the reflective wall initiates a significant effect on the extremal statistics. $P(M|x_0,t)$ can display a single maximum at an intermediate level of $M$, unlike free Brownian motion where $P(M|x_0,t)$ decays monotonically with $M$. In the long-time limit, the expectation $\langle M \rangle$ exhibits the same scaling with $t$ as the free free Brownian motion,  $\langle M(x_0,t) \rangle \sim t^{1/2}$, but the prefactor $\sqrt{\pi D }$ is $\pi/2$ times of free Brownian motion starting from the origin. Regarding to the distribution of $t_m$, $t_m$ has a higher probability occurring near $t$, and thus becomes asymmetric with respect to $t_m=t/2$. The ratio $\langle t_m \rangle/t$ increases monotonically from 1/2 as $t$ increases, and approaches another constant $(1+2G)/4 \approx0.708$ in the limit of $t \gg t_d$, where $G$ is the Catalan's constant.  Moreover, we investigate the statistics of the last time $t_\ell$ visited the initial position $x_0$ and the cumulative time $t_o$ spent above $x_0$. We exactly compute the expectations of $t_\ell$ and $t_o$. Interestingly, $\langle t_\ell \rangle/t$ is a nonmonotonic function of $t$, and a maximal $\langle t_\ell \rangle/t$ occurs at $t \approx 3.1216 x_0^2/D$.

It would be worthy to explore two possible extensions of the present study in the future. These include the study of EVS of Brownian motions in high dimensions in which the boundaries are set to be reflective and otherwise opening. For example, a planar  Brownian motion confined in a wedge geometry with reflective boundaries. Another interesting extension is to investigate the EVS of active Brownian motions in the presence of reflective boundary, such as run-and-tumble motions in one dimension confined in a half-axis by a reflective wall. It would be interesting to explore the effect of nonequilibrium noise on the EVS of Brownian motions.

\begin{acknowledgments}
This work was supported by the National Natural Science Foundation of China (Grants No. 11875069) and the Key Scientific Research Fund of Anhui Provincial Education Department (Grants No. 2023AH050116).
\end{acknowledgments}

\appendix
\section{The expectation $\langle x(t) \rangle$ of the displacement $x(t)$ with a reflective wall}\label{appen1}
We solve the diffusion equation Eq.(\ref{eq2.2.1}) in the Laplace domain, subject to a reflective boundary at $x=0$, ${\tilde G'} ( {0,s|{x_0}} ) = 0$. The boundary at $x=+\infty$ requires ${\tilde G} ( {+\infty,s|{x_0}} ) = 0$. On the other hand, the continuous condition in Eq.(\ref{eq2.6}) and the matching condition in Eq.(\ref{eq2.6}) are both satisfied. As done in Sec\ref{sec3b}, the solution is 
\begin{eqnarray}\label{eqa1.1}
\tilde G\left( {x,s|{x_0}} \right) = \frac{{{e^{ - \alpha \left( {x + {x_0}} \right)}} + {e^{\alpha \left( {{x_ < } - {x_ > }} \right)}}}}{{2\alpha D}}
\end{eqnarray}
where $\alpha=\sqrt{s/D}$, $x_<=\min(x,x_0)$ and $x_>=\max(x,x_0)$. Multiplying Eq.(\ref{eqa1.1}) by $x$ and then integrating over $x$ from 0 to $\infty$, we have 
\begin{eqnarray}\label{eqa1.2}
\langle {x( s )} \rangle  = \int_0^\infty  {dxx\tilde G\left( {x,s|{x_0}} \right)}=\frac{{{e^{ - \alpha {x_0}}} + \alpha {x_0}}}{{{\alpha ^3}D}}
\end{eqnarray}
By Laplace transform inversion for Eq.(\ref{eqa1.2}), we obtain the expectation $\langle x(t) \rangle$ of the displacement $x(t)$, 
\begin{eqnarray}\label{eqa1.3}
\langle {x( t )} \rangle  = 2{e^{ - \frac{{x_0^2}}{{4Dt}}}}\sqrt {\frac{{Dt}}{\pi }}  + {x_0}{\rm{erf}}\left( {\frac{{{x_0}}}{{2\sqrt {Dt} }}} \right)
\end{eqnarray}
where ${\rm{erf}}(x)$ is the error function. Considering the short-time limit and the long-time limit, we have
\begin{eqnarray}\label{eqa1.4}
\langle x(t) \rangle= \left\{ \begin{array}{cc}
x_0,   & t \ll t_d,  \\
2\sqrt {\frac{{Dt}}{\pi }}  + \frac{{x_0^2}}{{2\sqrt {\pi Dt} }}, &   t \gg t_d. \\ 
\end{array}  \right. 
\end{eqnarray}


\end{document}